\documentclass[reprint,showpacs,twocolumn,superscriptaddress,aps]{revtex4-2}
\usepackage[utf8]{inputenc}
\usepackage[american,british]{babel}
\usepackage[T1]{fontenc}
\usepackage[pdftex]{graphicx}  
\usepackage{graphicx}
\usepackage{xcolor}
\usepackage{mathtools}
\usepackage{bbm}
\usepackage{subfigure}
\usepackage{dcolumn}
\usepackage{braket}
\usepackage{bm}
\usepackage{amsmath,amsthm,amssymb}
\usepackage{color}
\usepackage{verbatim}
\usepackage{comment}

\definecolor{Red}{HTML}{E53E30}  % "Pantone 179"
\definecolor{Green}{HTML}{00AD69}  % "Pantone 3405"
\definecolor{Blue}{HTML}{2171b5}
\definecolor{Purple}{HTML}{652F6C}  % "Pantone 520"
\definecolor{forestgreen}{rgb}{0.1, 0.6, 0.2}

\newcommand{\eq}[1]{\begin{equation}#1\end{equation}}

\usepackage[colorlinks,citecolor=blue,linkcolor=blue,urlcolor=blue,hyperindex]{hyperref}

\usepackage[normalem]{ulem}

\usepackage[nameinlink,noabbrev]{cleveref}
\makeatletter
\def\@seccntformat#1{\@ifundefined{#1@cntformat}%
   {\csname the#1\endcsname\quad}%      default
   {\csname #1@cntformat\endcsname}%    enable individual control
}
\makeatother
\DeclareMathAlphabet\mathbfcal{OMS}{cmsy}{b}{n}
\usepackage{dsfont}
\usepackage{physics}
\begin{document} 

\title{The quantum Mpemba effect in free-fermionic mixed states}

\author{Filiberto Ares}

\affiliation{SISSA and INFN, via Bonomea 265, 34136 Trieste, Italy }

\author{Vittorio Vitale}
\affiliation{Univ.  Grenoble Alpes, CNRS, LPMMC, 38000 Grenoble, France}

\author{Sara Murciano}
\affiliation{Department of Physics and Institute for Quantum Information and Matter, California Institute of Technology, Pasadena, CA 91125, USA}

\affiliation{Walter Burke Institute for Theoretical Physics, California Institute of Technology, Pasadena, CA 91125, USA}

\date{\today}

\begin{abstract}

Recently, a novel probe to study symmetry breaking, known as entanglement asymmetry, has emerged and has been utilized to explore how symmetry is dynamically restored following quantum quenches. Interestingly, it has been shown that, in certain scenarios, greater initial symmetry breaking leads to faster restoration, akin to a quantum Mpemba effect. This study focuses on investigating the effect of mixed initial states and non-unitary dynamics on symmetry restoration. The \textit{mixedness} of a state can arise from different sources. We consider dephasing or dissipative processes affecting initial pure states or unitary dynamics of initially thermal states. In the former case, the stationary state after the quench is independent of the initial configuration, resembling the phenomenology of the classical Mpemba effect. Investigating the XY spin chain model, through a combination of analytical calculations and numerical simulations, we identify the conditions for the occurrence of the quantum Mpemba effect. It turns out that this phenomenon still occurs in the presence of dissipation or at finite temperature, even though it will be eventually suppressed as the state becomes more mixed.

\end{abstract}

\maketitle

%\tableofcontents
\section{Introduction}

A question barely considered in the literature is how an initially broken symmetry in a many-body quantum system evolves in time under a dynamics that preserves it. In a very recent work~\cite{amc-23}, 
this problem has been examined for a quantum quench: a spin-$1/2$ chain is initialized in a non-equilibrium state that breaks the $U(1)$ particle number symmetry and is let evolve under the unitary dynamics described by the Hamiltonian of the XX spin chain. Since the chain undergoes a unitary evolution, it never 
relaxes globally and the symmetry is not restored in the whole system. However, any portion of it does relax in the thermodynamic limit into a stationary state that respects the symmetry. A surprising finding is that the restoration of it may be faster for those initial states that break it more. This phenomenon can be seen as a quantum version of the Mpemba effect~\cite{mpemba-69}: the more a system is out of equilibrium, the faster it may relax. Although the Mpemba effect was for a long time viewed as a mere curiosity, today it is acknowledged as a genuine out-of-equilibrium phenomenon that can occur in many physical systems~\cite{ahn16, hu18, chaddah10, greaney11, lasanta17, keller18}, and is nowadays the subject of numerous papers, prompted by recent theoretical developments~\cite{lr-17, krhv-19, wv-22, tyr-23, wbv-23, bwv-23, brp-23, bp-24} and its observation in experimental controlled setups~\cite{kb-20, kcb-22}.   

After Ref.~\cite{amc-23}, the quantum Mpemba effect (QMPE) has been investigated in free fermionic systems, such as the XY spin chain~\cite{makc-23}, long-range Kitaev chains~\cite{carc-24}, or two-dimensional superconductors~\cite{yac-24}, and in generic one-dimensional interacting integrable systems~\cite{rkacmb-23, bkccr-23}, where the mechanisms and the criteria for its occurrence have been established. This phenomenon has been also reported in chaotic random unitary circuits~\cite{liu_mpemba_circuit-24}. In addition, it has been observed in experiments with an ion-trap quantum simulator that mimics the dynamics of a long-range XX spin chain~\cite{joshi-24}. In parallel, several theoretical~\cite{quantum1, quantum2, quantum3, quantum4, quantum5, quantum6, cth-23-2, spc-24, mczg-24} and experimental~\cite{shapira-24, zhang-exp-24} works have explored the occurrence of other Mpemba effects in open few-body quantum systems connected to thermal reservoirs or undergoing different non-unitary dynamics. 

A key ingredient of the setups considered to observe the symmetry restoration is that the full system is always initially prepared in a pure state and it is isolated from the environment, evolving unitarily after the quench. It is thus natural to wonder how this phenomenon is affected when the state does not remain pure during the whole dynamics. In realistic conditions, quantum systems are indeed described by mixed states as they are subject to external noise or decoherence. 
The effects of non-unitary dynamics on symmetry restoration have been analyzed in the experimental work~\cite{joshi-24} and in the theoretical work~\cite{cma-24}. In Ref.~\cite{joshi-24}, the external environment affects unavoidably the quantum state prepared in the experimental platform. In that setting, the main source of noise is global dephasing, which randomly rotates the spins around the $z$ axis. It is found that the presence of global dephasing does not spoil the QMPE, which occurs even if the dephasing rate is arbitrarily large. Contrarily, if the system evolves only subject to the dephasing noise of the laboratory, neglecting the unitary dynamics, the symmetry is indeed restored but the QMPE is absent. The theoretical work~\cite{cma-24} examines a quantum quench of a spin chain initially prepared in a tilted N\'eel state and undergoing an evolution described by the XX Hamiltonian in the presence of global gain and loss dissipation. Remarkably, in the pure unitary dynamics case, the particle number symmetry is not restored because the system locally relaxes to a non-Abelian generalized Gibbs ensemble (GGE)~\cite{amvc-23}. However, when the chain is connected to the environment via gain and loss terms, the symmetry is locally restored and the QMPE can occur.

The goal of the present manuscript is to analyze the effects of both 
non-unitary time evolution and initial symmetry-breaking mixed states in the context of symmetry restoration and QMPE. We consider a paradigmatic model: the XY spin chain. This is the simplest but non-trivial instance of a system that explicitly breaks 
the $U(1)$ particle number symmetry and includes the case of the XX spin chain -- which respects it. As shown in Ref.~\cite{makc-23}, this symmetry is dynamically restored for quantum states initialized in the ground state of the XY spin chain Hamiltonian and evolving under the XX one. For different values of the initial couplings, the QMPE can occur. Here we will study this quench in the presence of global gain and loss dissipation and, separately, of local dephasing. We will also consider the same quench dynamics initializing the chain at a finite temperature and keeping it disconnected from any environment after the quench. This setup is particularly useful for investigating how the {\it mixedness} of the initial configurations affects the symmetry restoration.

\begin{figure*}[t!]
\centering
\includegraphics[width=0.92\linewidth]{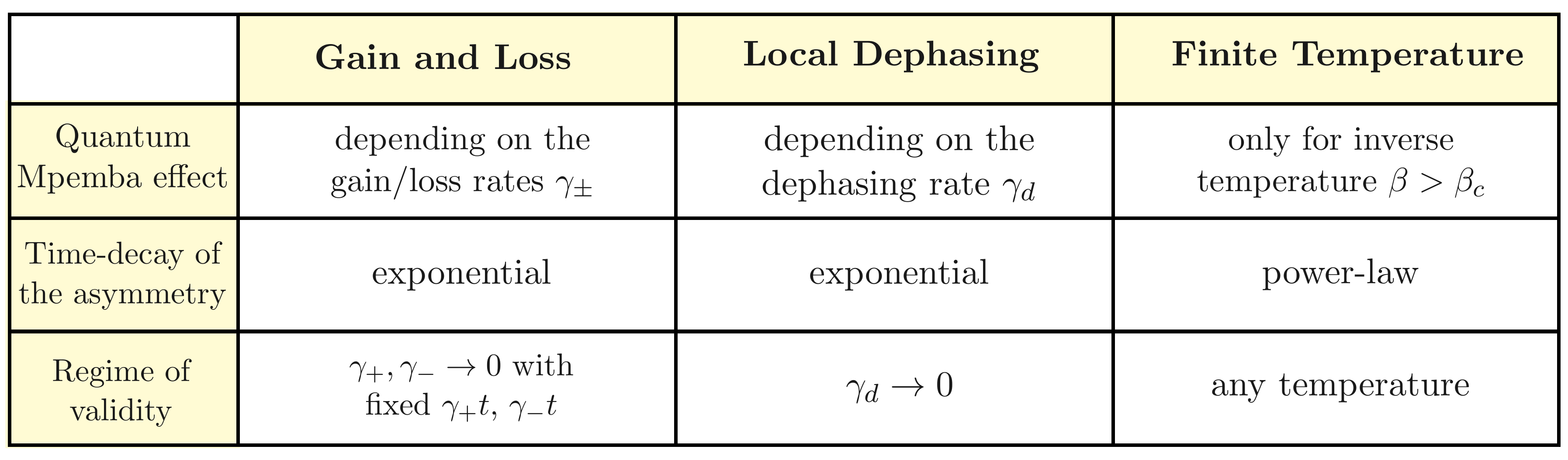}
\caption{Summary of the main findings of the manuscript. We study the entanglement asymmetry in three different mixed-state setups and we highlight whether the quantum Mpemba effect occurs or not, how fast the asymmetry decays to zero, and the regime of validity of our analytical predictions. }
\label{fig:table}
\end{figure*}

To probe the non-equilibrium dynamics of the symmetry and the QMPE, we employ the entanglement asymmetry~\cite{amc-23}.
This is a quantum information based observable that measures how much a symmetry is broken in a part of an extended quantum system. Apart from being useful to study $U(1)$ symmetries in quantum quenches, the entanglement asymmetry has been also applied to investigate discrete symmetries~\cite{fac-23, cm-23}, generic compact Lie groups in matrix product states~\cite{cv-23}, CFTs with non-topological defects~\cite{fadc-24, chen-23}, symmetry breaking Haar random states~\cite{ampc-23}, and confinement~\cite{Khor-23}. Here we calculate the entanglement asymmetry in the quantum quenches in the XY spin chain that we have mentioned; using the quasiparticle picture, we obtain
the exact analytic expression for the time evolution of the asymmetry
in the presence of gain and loss dissipation and when the system is initially prepared at a finite temperature. Moreover, we derive 
the conditions for the QMPE to occur in these cases. See Fig.~\ref{fig:table} for a summary of our results.

The paper is organized as follows. In Sec.~\ref{sec:review}, we 
introduce the entanglement asymmetry and we discuss how to calculate it from the charged moments of the reduced density matrix in the case 
of fermionic Gaussian states. We also review the main results obtained in Ref.~\cite{makc-23} for the entanglement asymmetry and the QMPE in a quantum quench from the ground state of an XY spin chain to the XX chain. In Sec.~\ref{sec:dissipation}, we study the effects on the entanglement asymmetry and the QMPE of gain and loss dissipation and local dephasing in this quench. In Sec.~\ref{sec:temperature}, we analyze the entanglement asymmetry in the XY spin chain at a finite temperature and then the time evolution after a quantum quench to the XX chain. In Sec.~\ref{sec:conclusions}, we draw our conclusions and discuss future prospects. We also include an appendix where we describe in detail the derivation of some results of the main text.

\section{Entanglement asymmetry and charged moments}\label{eq:review}

Let us start with an extended quantum system in a state described by a density matrix $\rho$, either mixed or pure, that can be divided into two spatial regions, $A$ and $B$. 
In our case, we will take as subsystem $A$ an interval of $\ell$ contiguous sites of the spin chain.
This is the geometry that we will consider throughout the manuscipt.
The state of the system in $A$ is given by the reduced density matrix $\rho_A=\mathrm{Tr}_B(\rho)$,
obtained by tracing out the degrees of freedom of the complementary subsystem $B$. We consider that the dynamics of the system has a global additive conserved charge, $Q=Q_A\otimes \mathds{1}_B+\mathds{1}_A\otimes Q_B$, that generates a $U(1)$ symmetry group. If $\rho$ respects this symmetry, then
$[\rho_A, Q_A]=0$ and $\rho_A$ displays a block-diagonal structure in the charge sectors of $Q_A$. In that case, a very active problem is to identify and compute the contributions to the entanglement of each symmetry sector~\cite{lr-14, gs-18, xavier, brc-19, mdgc-20, pbc-21, bcckr-23, mca-23, lukin-19, azses-20, neven-21, vitale2022symmetry, rvm-22}. On the other hand, if the
state $\rho_A$ breaks the symmetry, then $[\rho_A, Q_A]\neq 0$ and $\rho_A$ is not block-diagonal. To measure how much $\rho_A$ breaks the symmetry, we can use the R\'enyi entanglement asymmetry, defined as~\cite{amc-23}
\begin{eqnarray}\label{eq:renyiEA}
     \Delta S_A^{(n)}(\rho)=S^{(n)}(\rho_{A, Q})-S^{(n)}(\rho_A),
\end{eqnarray}
where
\begin{eqnarray}\label{eq:renyiE}
     S^{(n)}(\rho)=\frac{1}{1-n}\mathrm{log}\mathrm{Tr}(\rho^n),
\end{eqnarray}
are the R\'enyi entropies. The matrix $\rho_{A,Q}$ is obtained from $\rho_A$ as $\rho_{A,Q}=\sum_{q\in\mathbb{Z}}\Pi_q\rho_A\Pi_q$, where $\Pi_q$ is 
the projector onto the eigenspace of $Q_A$ with charge 
$q$. Thus $\rho_{A, Q}$ is block-diagonal in the 
eigenbasis of $Q_A$ and, therefore, it is symmetric, i.e. $[\rho_{A, Q}, Q_A]=0$. In the limit $n\to 1$, one finds~\cite{hlw-94, cc-04}
\begin{equation}\label{eq:def}
 \Delta S_{A}=S(\rho_{A,Q})-S(\rho_A),
\end{equation}
where 
\begin{equation}\label{eq:vn}
    S(\rho)=-{\rm Tr}(\rho\log \rho)
\end{equation}
is the von Neumann entropy associated with the density matrix $\rho$. The von Neumann entanglement asymmetry \eqref{eq:def} can be further rewritten as the relative entropy between the symmetry-broken reduced density matrix, $\rho_A$, and its symmetrized version, $\rho_{A,Q}$ \cite{Ma2022}.

%As in previous works, to calculate the asymmetry we can exploit the \textit{replica trick}, and thus we resort to the R\'enyi entropies,

%With this in mind, the definition of Eq.~\eqref{eq:def} can be extended by replacing the von Neumann entropy $S(\rho)$ with the R\'enyi entropies $S^{(n)}(\rho)$,
%\begin{equation}\label{eq:replicatrick}
% \Delta S_{A}^{(n)}=S^{(n)}(\rho_{A, Q})-S^{(n)}(\rho_A).
% \end{equation}

As already discussed in Ref.~\cite{amc-23}, both the R\'enyi~\eqref{eq:renyiEA} and von Neumann~\eqref{eq:def}  entanglement asymmetries satisfy two fundamental properties as measures of symmetry breaking:
\emph{(i)} they are non-negative, $\Delta S_{A}^{(n)}\geq 0$, \emph{(ii)} they vanish, $\Delta S_{A}^{(n)}=0$, iff the state of subsystem $A$ respects the symmetry associated to $Q$, namely when $[\rho_A, Q_A]=0$.

However, in this manuscript, we will mainly compute the asymmetry using Eq.~\eqref{eq:renyiEA}: The main advantage of the R\'enyi entanglement asymmetry is that it 
is easier to calculate for integer $n\geq 2$, and Eq.~\eqref{eq:def} can be recovered by doing the analytic continuation in $n$ and 
taking the limit $\lim_{n\to 1}\Delta S_{A}^{(n)}=\Delta S_{A}$. 
Moreover, for integer $n\geq 2$, it can be experimentally accessed~\cite{joshi-24}
via randomized measurements~\cite{elben2023,brydges2019probing,elben2020mixed,vitale2022symmetry,rvm-22,satzinger2021,Yu2021,rozon2024}. For these reasons, in the following sections we will focus on the computation of the R\'enyi entanglement asymmetry, computing the replica limit $n\to 1$ only when it is possible.

\subsection{Charged moments}

The calculation of the R\'enyi entanglement asymmetries~\eqref{eq:renyiEA} for integer $n\geq 2$ is usually simpler because they can be expressed in terms of the charged moments of $\rho_A$, which can be explicitly computed in several settings, e.g. for fermionic Gaussian states~\cite{makc-23, rkacmb-23}, in CFTs~\cite{fadc-24, chen-23}, or for Haar random states~\cite{ampc-23}. Using the Fourier representation of the projectors $\Pi_q$, $\rho_{A, Q}$ can be rewritten as
\begin{equation}
 \rho_{A, Q}=\int_{-\pi}^\pi \frac{{\rm d}\alpha}{2\pi}e^{-i\alpha Q_A}\rho_A e^{i\alpha Q_A}.
\end{equation}
Therefore, its moments are given by
\begin{equation}\label{eq:FT}
 \mathrm{Tr}(\rho_{A, Q}^n)=\int_{-\pi}^\pi \frac{{\rm d}\alpha_1\dots{\rm d}\alpha_n}{(2\pi)^n} Z_n(\boldsymbol{\alpha}),
 \end{equation}
where $\boldsymbol{\alpha}=\{\alpha_1,\dots,\alpha_n\}$ and $Z_n(\boldsymbol{\alpha})$ are the (generalized) charged moments
\begin{equation}\label{eq:Znalpha}
 Z_n(\boldsymbol{\alpha})=
 \mathrm{Tr}\left[\prod_{j=1}^n\rho_A e^{i\alpha_{j,j+1}Q_A}\right],
\end{equation}
with $\alpha_{ij}\equiv\alpha_i-\alpha_j$ and $\alpha_{n+1}=\alpha_1$. We show in the next section how $Z_n(\boldsymbol{\alpha})$ can be efficiently computed for fermionic Gaussian states. 

\subsection{Unitary dynamics at zero temperature}\label{sec:review}

In order to analyze the results that we find in the manuscript, it is useful to recall the behavior of the entanglement asymmetry in the quantum quench from the XY to the XX spin chain studied in~\cite{makc-23}. In that work, the chain is initialized in the ground state $\ket{\Psi(0)}$ of the XY spin chain of $N$ sites
\begin{equation}\label{eq:Ham_XY}
H_{\rm XY}=-\frac{1}{4}\sum_{j=1}^{N}[(1+\delta)\sigma^x_j \sigma^x_{j+1}+(1-\delta)\sigma^y_j \sigma^y_{j+1} +2h \sigma_j^z],
\end{equation}
which breaks the $U(1)$ symmetry generated by the transverse magnetization $Q=\frac{1}{2}\sum_j\sigma^z_j$. Here $\sigma_j^{\beta}$ are the Pauli matrices at the site $j$, $\delta\neq 0$ is the anisotropy parameter breaking the $U(1)$ symmetry and $h$ is the value of the external transverse magnetic field. The $N$ sites constitute the full system, $A\cup B$, and we will take as subsystem $A$ an interval of $\ell$ contiguous spins. The Hamiltonian in Eq.~\eqref{eq:Ham_XY} can be written in terms of the fermionic operators $\boldsymbol{a}_j=(a_j^\dagger, a_j)$ via a Jordan-Wigner transformation, which yields
\begin{equation}\label{eq:XYfermi}
H_{\rm XY}=-\frac{1}{2}\sum_{j=1}^{N}\left(a^{\dagger}_j a_{j+1}+\delta a^{\dagger}_ja^{\dagger}_{j+1} +\mathrm{h.c.}+2h a^{\dagger}_ja_j\right),
\end{equation}
where we have neglected the boundary terms as we will always consider $N\gg \ell$.

The ground state $\ket{\Psi(0)}$ is then let evolved with the XX spin chain Hamiltonian $H_{\rm XX}$, which corresponds to Eq.~\eqref{eq:XYfermi} with $\delta=h=0$, 
\begin{equation}\label{eq:quench}
 \ket{\Psi(t)}= e^{-i tH_{\rm XX}}\ket{\Psi(0)}.
\end{equation}
The XX spin chain Hamiltonian respects the $U(1)$ symmetry because $[H_{\rm XX},Q]=0$. Since the full system follows a unitary evolution, it does not globally relax to a stationary state and the symmetry is never globally restored after the quench. On the other hand, in the thermodynamic limit $N\to\infty$, the state $\rho_A(t)$ of the subsystem $A$ for a finite length $\ell$ does converges to a stationary state, described by a generalized Gibbs ensemble, that commutes with $Q_A$~\cite{amc-23}. Therefore, the time evolution~\eqref{eq:quench} leads to a dynamical restoration of the $U(1)$ symmetry in the subsystem $A$. 

The reduced density matrix $\rho_A(t)$ corresponding to the time-evolved state~\eqref{eq:quench} is Gaussian in terms of the fermionic variables $\boldsymbol{a}_j$. Thus the charged moments $Z_n(\boldsymbol{\alpha})$ defined in Eq.~\eqref{eq:Znalpha} can be expressed through the two-point correlation function \cite{peschel2003, peschel2009}
\begin{equation}\label{eq:def_corr}\Gamma_{jj'}=2\mathrm{Tr}\left[\rho_A\boldsymbol{a}_j^\dagger
 \boldsymbol{a}_{j'}\right]-\delta_{jj'},
\end{equation}
with $j,j'\in A$. For the quench~\eqref{eq:quench}, this matrix can be explicitly computed and, in the thermodynamic limit $N\to \infty$, it reads
\begin{equation}\label{eq:Gammat0}
\Gamma_{jj'}=\int_{0}^{2\pi}\frac{{\rm d}k}{2\pi}\mathcal{G}_0(k,t)e^{-ik(j-j')}, \quad j,j'=1,\dots,\ell,
\end{equation}
with $\ell$ the size of the subsystem $A$.
This expression shows that $\Gamma$ is a block Toeplitz matrix with  symbol $\mathcal{G}_0(k,t)$ the $2\times 2$ matrix
\begin{equation}
  \mathcal{G}_0(k,t)=  \cos \Delta_k \sigma^z+\sin \Delta_k \sigma^y e^{2it \epsilon_k \sigma^z},
\end{equation}
where
\begin{equation}
\begin{split}\label{eq:cs}
 \cos\Delta_k&=\frac{h-\cos k}{\sqrt{(h-\cos k)^2+\delta^2\sin^2 k}}, \\
  \sin\Delta_k&=\frac{\delta \sin k}{\sqrt{(h-\cos k)^2+\delta^2\sin^2 k}},
 \end{split}
\end{equation}
and $\epsilon_k=-\cos(k)$ is the single-particle dispersion relation of the XX Hamiltonian. In terms of the two-point correlation matrix $\Gamma$, the charged moments are given by the determinant~\cite{amc-23, amvc-23}
\begin{equation}\label{eq:numerics}
  Z_n(\boldsymbol{\alpha})=\sqrt{\det\left[\left(\frac{I-\Gamma}{2}\right)^n
  \left(I+\prod_{j=1}^n W_j\right)\right]},
\end{equation}
with $W_j=(I+\Gamma)(I-\Gamma)^{-1}e^{i\alpha_{j,j+1} n_A}$ and $n_A$ is a diagonal matrix with $(n_A)_{2j,2j}=1$, $(n_A)_{2j-1,2j-1}=-1$, $j=1, \dots, \ell$. 

\begin{figure}[t]
\includegraphics[width=0.45\textwidth]{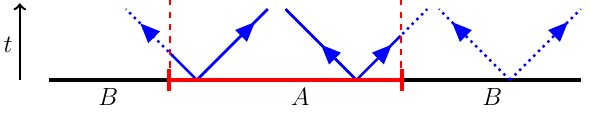}
\caption{Schematic representation of the quasiparticle picture. In the quench, pairs of entangled quasiparticles propagate ballistically with opposite momentum. In our case, each pair also forms a Cooper pair, which is responsible for breaking the particle-number symmetry. The Cooper pairs with both excitations inside the subsystem $A$ contribute to the asymmetry (solid blue arrows). When one of the excitations leaves $A$, the pair stops contributing to the asymmetry (dashed blue arrows).}
\label{fig:qpp}
\end{figure}

Combining~\eqref{eq:numerics} and some properties of the determinants of block Toeplitz matrices, the full time evolution of the charged moments 
can be exactly obtained in the hydrodynamic limit $t,\ell\to\infty$ with $\zeta=t/\ell$ fixed. 
In that regime, they read~\cite{makc-23}
\begin{equation}\label{eq:charged_moments_ev_unitary}
Z_n(\boldsymbol{\alpha}, t)= Z_n(\boldsymbol{0}, t)e^{B_n(\boldsymbol{\alpha, \zeta)}\ell},
\end{equation}
where
\begin{equation}\label{eq:B_n_unitary}
B_n(\boldsymbol{\alpha}, \zeta)=\int_{0}^{2\pi}\frac{{\rm d}k}{4\pi}
x_k(\zeta)\sum_{j=1}^n f(\cos\Delta_k, \alpha_{j, j+1}),
\end{equation}
with $x_k(\zeta)=1-\min(2|v_k|\zeta, 1)$, $v_k=\epsilon'_k$, and $f(\lambda, \alpha)=\log(i\lambda\sin\alpha+\cos\alpha)$.
This expression can be interpreted using the quasiparticle picture 
of quantum quenches~\cite{cc-05, ac-17, ac-18}. This is a semiclassical picture that explains the non-equilibrium dynamics of the entanglement entropy $S(\rho_A)$ and many other quantities after a quench in integrable systems, see the pedagogical reviews~\cite{c-18, c-20}. According to it, in the quench~\eqref{eq:quench}, pairs of 
entangled quasiparticles are produced and spread ballistically with 
opposite momentum and velocities $v_{\pm k}$, as we illustrate in Fig.~\ref{fig:qpp}. The key tenet of the dynamics of the charged moments (and of the entanglement entropy) is how the members of the entangled pair are shared between the subsystem $A$ and the rest. In the standard quasiparticle picture for the entanglement entropy~\cite{cc-05, ac-17, ac-18}, only the pairs shared between $A$ and $B$ contribute to the entropy. In the entanglement asymmetry, instead, only  the configurations in which the two entangled quasiparticles are in the subsystem $A$ contribute to the ratio $Z_n(\boldsymbol{\alpha}, t) /Z_n(\boldsymbol{0}, t)$. In Eq.~\eqref{eq:B_n_unitary}, the term $ x_k(\zeta)$ accounts for the number of entangled pairs of momentum $\pm k$ contained in the subsystem $A$ at time $t$ while the rest of the integrand represents their contribution to $Z_n(\boldsymbol{\alpha}, t) /Z_n(0, t)$. These are the pairs represented by solid arrows in Fig.~\ref{fig:qpp}. Since the quasiparticles forming the pair have opposite velocities, at long times they cannot be simultaneously in $A$, as we can see in the figure, stop contributing to the asymmetry. These are represented by dashed arrows in Fig.~\ref{fig:qpp}. In the thermodynamic limit $N\to\infty$, the number of pairs with both excitations inside $A$ tends to zero when $t\to\infty$; therefore, the asymmetry vanishes and the $U(1)$ symmetry is restored in $A$.
In fact, if we go back to Eq.~\eqref{eq:B_n_unitary}, in the limit $t\to\infty$, we have that $x_k(\zeta)\to0$, implying that $Z_n(\boldsymbol{\alpha}, t )\to Z_n(0, t)$
and, consequently, $\Delta S_A^{(n)}(t)\to 0$.
This picture has been extended to quantum quenches that generate entangled multiplets~\cite{bastianello2018, bastianello2020, caceffo2023} of arbitrary size in Ref.~\cite{cma-24}; for example, initial configurations that break translational invariance, in which the symmetry may be not restored~\cite{amvc-23}.

From the charged moments~\eqref{eq:charged_moments_ev_unitary}, we can compute the time evolution of the R\'enyi entanglement asymmetries by inserting them in Eq.~\eqref{eq:FT}. In particular, at time $t=0$, they grow logarithmically with the subsystem size,
\begin{equation}
    \label{eq:asym_XY_t0}
    %\begin{split}
    \Delta S_A^{(n)}(\delta,h,t=0)=\frac{1}{2} \log \ell + \frac{1}{2}\log\frac{\pi g(\delta, h)n^{\frac{1}{n-1}}}{4}+ O(\ell^{-1}).
    %\end{split}
\end{equation}
In this expression, the coefficient of the $\log\ell$ term is related to the dimension of the symmetry group~\cite{cv-23} and
\begin{equation}\label{eq:g_gs}
g(\delta, h)=\int_0^{2\pi} \frac{{\rm d}k}{2\pi} \sin^2 \Delta_k(\delta,h).
\end{equation}
As noted in Ref.~\cite{makc-23}, $\sin\Delta_k$ is the mode occupation of 
Cooper pairs in the ground state of the XY spin chain~\eqref{eq:Ham_XY}. Therefore,
according to Eq.~\eqref{eq:asym_XY_t0}, the initial entanglement asymmetry increases with the number of Cooper pairs that the chain contains in the ground state. This a quite natural result since the pairing term in the Hamiltonian~\eqref{eq:XYfermi} that breaks the $U(1)$ particle number symmetry induces the creation of Cooper pairs.

On the other hand, at times $t\gg \ell$, the R\'enyi entanglement asymmetries behave as
\begin{equation}
    \label{eq:asym_XY_tinf}
    %\begin{split}
    \Delta S_A^{(n)}(\delta,h,t)\simeq \frac{n\ell}{1-n} \int_0^{2\pi} \frac{{\rm d}k}{16\pi} x_k(\zeta)\sin^2 \Delta_k (\delta,h).
     %\end{split}
\end{equation}
As a consequence, at late times after the quench, the entanglement asymmetry is
governed by the slowest Cooper pairs that are still inside the subsystem $A$.

Now we can take two initial ground states, $\ket{\Psi_1(0)}$ and $\ket{\Psi_2(0)}$, corresponding to different couplings, $(\delta_1, h_1)$ and $(\delta_2, h_2)$, of the XY spin chain~\eqref{eq:Ham_XY}, for which 
\begin{equation}\label{eq:condition_mpemba_t_0}
\Delta S_A^{(n)}(\delta_1, h_1, t=0)>\Delta S_A^{(n)}(\delta_2, h_2, t=0).\end{equation}
In other words, $\ket{\Psi_1(0)}$ breaks more the $U(1)$ particle-number symmetry than $\ket{\Psi_2(0)}$. It can occur that there is a time $t_I$ after which the relation~\eqref{eq:condition_mpemba_t_0} is inverted,
\begin{equation}\label{eq:condition_mpemba_t_I}
\Delta S_A^{(n)}(\delta_1, h_1, t)<\Delta S_A^{(n)}(\delta_2, h_2, t), \quad t>t_I.
\end{equation}
This means that the entanglement asymmetries of $\ket{\Psi_1(0)}$ and $\ket{\Psi_2(0)}$ intersect at a certain finite time and the symmetry is restored faster for $\ket{\Psi_1(0)}$, indicating that there is quantum Mpemba effect. Although for some fine tuned initial states the entanglement asymmetries can exhibit multiple intersections at finite time~\cite{carc-24}, here we will assume that only one intersection can occur.

Eqs.~\eqref{eq:condition_mpemba_t_0} and \eqref{eq:condition_mpemba_t_I} are the necessary and sufficient conditions for the occurrence of this phenomenon. Using~\eqref{eq:asym_XY_t0} and~\eqref{eq:asym_XY_tinf} respectively, we can 
re-express them in terms of the density of Cooper pairs in the initial states $\ket{\Psi_1(0)}$ and $\ket{\Psi_2(0)}$~\cite{makc-23}
\begin{equation}
    \label{eq:mpemba_XY}
    \begin{cases}
        \displaystyle \int_{0}^{2\pi}\frac{{\rm d}k}{2\pi}\sin^2\Delta_k(\delta_1,h_1)>\int_{0}^{2\pi}\frac{{\rm d}k}{2\pi}\sin^2\Delta_k(\delta_2,h_2),\\
    \displaystyle   \int_{-k^*_{\zeta}}^{k^*_{\zeta}} \frac{{\rm d}k}{2\pi} \Upsilon_k(\delta_1,h_1)<\int_{-k^*_{\zeta}}^{k^*_{\zeta}} \frac{{\rm d}k}{2\pi} \Upsilon_k(\delta_2,h_2),\, \text{for } t>t_I,
    \end{cases}
\end{equation}
where
\begin{equation}\label{eq:Upsilon_gs}
\Upsilon_k(\delta,h)=\sin^2\Delta_k(\delta,h)+\sin^2\Delta_{k+\pi}(\delta,h)
\end{equation}
and $k^*_{\zeta}=\arcsin(1/(2\zeta))$.
These two conditions show that the QMPE can be observed when the less asymmetric initial state, which contains a smaller number of Cooper pairs, has instead a larger density of slowest Cooper pairs,  those with momenta around $k=0$ and $k=\pi$. 

As a final remark, we stress that even though the symmetry is not globally restored, locally the system relaxes to a symmetric stationary state. In order to get that state, we have to consider first an infinite system, $N\to\infty$, and then take the limit $t\to \infty$, and only at the end the limit of a large subsystem $\ell\to\infty$.  Since the limits $\ell\to \infty$ and $t\to\infty$ do not commute, if we first consider $\ell\to\infty$, the asymmetry diverges and the symmetry will never be restored at large times. The quasiparticle picture prediction~\eqref{eq:charged_moments_ev_unitary} corresponds to the leading order behavior of the charged moments when we simultaneously take $t,\ell\to\infty$ keeping its ratio fixed, neglecting the subleading corrections in $\ell$, which go to zero when $\ell\to\infty$~\cite{ac-17, ac-18}.

\section{Global quench under dissipative evolution}\label{sec:dissipation}

After this brief review of the behavior of the entanglement asymmetry under a unitary time evolution, we ask what happens if in the quench~\eqref{eq:quench} there are dissipative effects and the evolution is non-unitary. In particular, we assume that the non-equilibrium dynamics of the total density matrix $\rho(t)$ is now described by the Lindblad equation \cite{petruccione2002the}
\begin{equation}
\label{eq:lind}
\begin{split}
\frac{{\rm d}\rho(t)}{{\rm d}t}=&\mathcal{L}(\rho(t))=-i[H_{\rm XX},\rho(t)]\\&+\sum_{j=1}^N\sum_{\alpha=\pm}\left(L_{j,\alpha}\rho(t) 
L_{j,\alpha}^\dagger-\frac{1}{2}\left\{L_{j,\alpha}^\dagger L_{j,\alpha},\rho(t)\right\}\right)\, .
\end{split}
\end{equation}
In this formula, $H_{\rm XX}$ is the Hamiltonian of the XX spin chain and the dissipation is encoded in the Lindblad jump operators $L_{j,\alpha}$, which in our case will be of the form $L_{j,-}=\sqrt{\gamma_-}a_j$, $L_{j,+}=\sqrt{\gamma_+}a_j^\dagger$, and $L_{j,z}=\sqrt{\gamma_d}a_j^\dagger a_j$. The first two correspond to the creation and annihilation
of fermions in all sites at rates $\gamma_+$ and $\gamma_-$, respectively, and they preserve the Gaussianity of $\rho(t)$ during the dynamics; therefore, we can employ Eq.~\eqref{eq:numerics} to study the entanglement asymmetry. 
We remark that, recently, it has been found that the entanglement dynamics in systems modeled by quadratic fermionic and bosonic Lindblad master equations is described by the quasiparticle picture~\cite{alba2021spreading,carollo2022dissipative,alba2022logarithmic,alba2022hydrodynamics}.
The last jump operator describes a dephasing with rate $\gamma_d$. In this case, the Liouvillian in Eq.~\eqref{eq:lind} is no longer quadratic and, in principle, we cannot exploit the machinery of the previous sections, only valid for Gaussian states. However, following the approach applied in Ref.~\cite{alba2021spreading} to study the entanglement entropy and the negativity in this scenario, we will neglect the deviations from the Gaussian behavior, and we will compute the R\'enyi entanglement asymmetry with the two-point correlation matrix using the formula~\eqref{eq:numerics}. As we will see, the time evolution of these correlation functions can be obtained efficiently, at least numerically. 
Let us first study the case of a dissipative evolution with gain and loss but no dephasing.
In the second part of this section, we investigate the effects of dephasing with no gain and loss terms.

\subsection{Gain and loss dissipation}\label{sub:gainloss}
\subsubsection{Charged moments}

In the presence of only linear gain and loss terms, the time-evolved
total density matrix $\rho(t)$ is Gaussian and, therefore, it is univocally determined by the two-point fermionic correlation functions~\eqref{eq:def_corr}. Using the results of Refs.~\cite{alba2022hydrodynamics,prosen2008}, one can find that, in this kind of quench, these correlations are of the form~\eqref{eq:Gammat0} with the symbol $\mathcal{G}_0(k,t)$ replaced by
\begin{equation}
\label{eq:symbol_dissipation}
\begin{split}
  \mathcal{G}(k,t)=  &\lambda(t)
  (\cos \Delta_k \sigma^z+\sin \Delta_k \sigma^y e^{2it \epsilon_k \sigma^z})
  \\-&\frac{\gamma_--\gamma_+}{\gamma_++\gamma_-}(1-\lambda(t))\sigma^z,
  \end{split}
\end{equation}
and $\lambda(t)=e^{-(\gamma_++\gamma_-) t}$.
The first step to study the entanglement asymmetry is to derive the time evolution of the charged moments $Z_n(\boldsymbol{\alpha}, t)$, defined in Eq.~\eqref{eq:Znalpha}, in the weakly-dissipative hydrodynamic limit $t, \ell \to \infty$, $\gamma_+,\gamma_- \to 0$ with fixed $t/\ell$, $\gamma_+ t$, $\gamma_-t$. As shown in Ref.~\cite{cma-24} in a quench to the XX spin chain from the N\'eel state with similar gain and loss terms, the quasiparticle picture for the charged moments discussed in Sec.~\ref{sec:review} remains valid in this regime but the dissipation modifies the contribution of the excitations to the charged moments.
As we show in Appendix~\ref{app:sub1}, the symbol~\eqref{eq:symbol_dissipation} is the starting point to derive the time evolution of the charged moments. In particular, in that appendix, we find that, in the weakly-dissipative hydrodynamic regime, they evolve as
\begin{equation}\label{eq:charged_m}
Z_n(\boldsymbol{\alpha}, t)= Z_n(\boldsymbol{0},t)e^{B_n(\boldsymbol{\alpha}, \zeta)\ell},
\end{equation}
where 
\begin{equation}\label{eq:B_n_dissipation}
 B_n(\boldsymbol{\alpha}, \zeta)=\int_0^{2\pi} \frac{{\rm d}k}{4\pi}x_k(\zeta)\log \left[\frac{\det \mathcal{M}_{\boldsymbol{\alpha}}^{(n)}(k, t\to0)}{\det\mathcal{M}_{\boldsymbol{0}}^{(n)}(k, t\to 0)}\right].
\end{equation}
%and
%\bigskip
%\begin{multline}\label{eq:neutral_mom_diss}
%Z_n(\boldsymbol{0},t)=\ell\int_0^{2\pi} \frac{{\rm d}k}{2\pi}\mathrm{min}(2\zeta |v_k|,1)h_n(n(k)) \\
%+\frac{\ell}{2}\int_{0}^{2\pi}\frac{{\rm d}k}{2\pi}x_k(\zeta)\log \det\mathcal{M}_{\boldsymbol{0}}^{(n)}(k, t\to 0),
% \end{multline}
%with   
%$h_n(x)=\log\left[x^n+(1-x)^n\right]$
%and 
%\begin{equation}\label{eq:occupation}
%\begin{split}
%    n(k, t)=\frac{1}{2}\left[1-\lambda(t)\cos \Delta_k  +\frac{\gamma_--\gamma_+}{\gamma_++\gamma_-}(1-\lambda(t))\right].
%\end{split}
%\end{equation}
This term depends on the $2\times 2$ matrix $\mathcal{M}_{\boldsymbol{\alpha}}^{(n)}(k, t)$
\begin{equation}\label{eq:M}
\mathcal{M}_{\boldsymbol{\alpha}}^{(n)}(k, t)=\left(\frac{I-\mathcal{G}(k,t)}{2}\right)^n
\left[I+\prod_{j=1}^n\frac{I+\mathcal{G}(k,t)}{I-\mathcal{G}(k,t)}e^{i\alpha_{j, j+1}\sigma^z}\right]
\end{equation}
and $\mathcal{G}(k, t\to 0)$, which is the limit $t\to0$ of Eq.~\eqref{eq:symbol_dissipation} keeping the terms $e^{-(\gamma_++\gamma_-)t}$, as they are constant in the weakly-dissipative regime,
\begin{equation}\label{eq:symbol_asymptotict0_0}
\begin{split}
\mathcal{G}(k,t\to 0)&=  \lambda(t)(\cos \Delta_k \sigma^z+\sin \Delta_k \sigma^y )\\&-\frac{\gamma_--\gamma_+}{\gamma_++\gamma_-}(1-\lambda(t))\sigma^z.
\end{split}
\end{equation}

We can compare this result with the one in Eq.~\eqref{eq:charged_moments_ev_unitary} for the case of 
a unitary quench with no dissipation. Both expressions are formally identical and the only difference resides in the integrand of the time dependent exponent $B_n(\boldsymbol{\alpha}, \zeta)$. Since the term $x_k(\zeta)$ remains untouched, only those entangled pairs with both particles inside the subsystem $A$  at time $t$ count but, as already announced, their contribution is altered by the dissipation. While for the pure unitary case it is determined by the two point correlations of the initial state, here it is encoded in the matrix~$\mathcal{M}_{\boldsymbol{\alpha}}^{(n)}(k, t\to0)$, which is given by the short-time two-point correlations~\eqref{eq:symbol_asymptotict0_0} in the weakly-dissipative limit. Unlike the case without dissipation, the contribution of the quasiparticle pairs is now time dependent. We note that, if we take $\gamma_{\pm}=0$ in Eq.~\eqref{eq:charged_m}, we recover the non-dissipative result~\eqref{eq:charged_moments_ev_unitary}. %Eq.~\eqref{eq:neutral_mom_diss} is the time evolution of the (neutral) moments of $\rho_A$ in the presence of linear gain and loss dissipation found in Ref.~\cite{alba2021spreading}. 

We now want to provide an explicit analytical expression for Eq.~\eqref{eq:B_n_dissipation}, which amounts to computing the early time behavior of the charged moments. Unfortunately, this is not an easy task except for some particular cases that we are going to examine. 
We start with $n=2$, for which a straightforward calculation of the determinant of the matrix $\mathcal{M}_{\boldsymbol{\alpha}}^{(n)}(k, t\to0)$ in Eq.~\eqref{eq:B_n_dissipation} leads to
\begin{widetext}
\begin{multline}\label{eq:gammapm}
  B_2(\alpha,\zeta)\sim \frac{1}{2} \int_0^{2\pi}\frac{{\rm d}k}{2\pi}
x_k(\zeta)\log \left[1-\sin^2\alpha\frac{\lambda^2(t)(\gamma_++\gamma_-)^4\sin^2\Delta_k}{(\gamma_-^2+\gamma_+^2)(1+\lambda^2(t))-\lambda(t)(\gamma_--\gamma_+)^2+(1-\lambda(t))(\gamma_-^2-\gamma_+^2)\cos^2\Delta_k}\right].
\end{multline}
\end{widetext}
For larger integer values of $n$, we can still find an expression of the charged moments using Eq.~\eqref{eq:B_n_dissipation}. However, the explicit formula becomes more and more cumbersome and it is not as compact as Eq.~\eqref{eq:gammapm} unless we require that gain and loss are balanced, $\gamma_+=\gamma_-$, and we take the large time regime, in which $\lambda(t)=e^{-2\gamma_+ t}\ll 1$. 
In that case, using the result in Eq.~\eqref{eq:B_n_dissipation}, we find that the leading order behavior in this regime is given by
\begin{equation}\label{eq:larget}
\begin{split}
   B_n(\boldsymbol{\alpha},\zeta)&\sim \frac{1}{2} \int_0^{2\pi}\frac{{\rm d}k}{2\pi}
x_k(\zeta)\\&\times \log\left[1-4 \lambda(t)^2\sin^2\Delta_k \sum_{i<j}\sin^2(\alpha_{ij}) \right].
\end{split}
\end{equation}
The main obstruction to finding a closed formula for the charged moments beyond the regime $\lambda(t)\ll 1$ is that, as also occurs for the balanced gain and loss case~\eqref{eq:larget}, the coefficient $B_n(\boldsymbol{\alpha}, \zeta)$ in Eq.~\eqref{eq:B_n_dissipation} does not admit a factorization in the replica space; that is, it cannot be written as
\begin{equation}\label{eq:fact_rep_space}
B_n(\boldsymbol{\alpha}, \zeta)=\sum_{j=1}^n B_1(\alpha_{jj+1},\zeta),
\end{equation}
as happens in the absence of dissipation, cf. Eq.~\eqref{eq:B_n_unitary}. 

%We observe that the result in Eq.~\eqref{eq:larget} agrees with the conjectured quasiparticle picture for quantum quenches generating entangled multiplets of arbitrary size proposed in Refs. \cite{bastianello2018,caceffo2023}.  The key tenet of the dynamics of the charged moments (and of the entanglement entropy) is how the members of the entangled multiplet are sharedbetween the subsystem $A$ and the rest.
%Here, the main difference with respect to the standard quasiparticle picture is that only configurations in which at
%least two entangled quasiparticles are in the subsystem contribute to the ratio $Z_n(\boldsymbol{\alpha}, t) /Z_n(0, t)$. If the quasiparticles forming the multiplet have different velocities (as it happens in the quench studied here), at long times any two
%quasiparticles cannot be simultaneously in $A$. By applying the same strategy used in \cite{cma-24}, one can prove that also in the presence of dissipation, when the entangled pairs are shared between $A$ and $B$, they do not contribute to $Z_n(\boldsymbol{\alpha}, t) /Z_n(\boldsymbol{0}, t)$ and the logarithmic term in Eq.~\eqref{eq:larget} comes from configurations where both entangled pairs are present in $A$.

\subsubsection{Entanglement asymmetry}

\begin{figure*}
\includegraphics[width=0.4\linewidth]{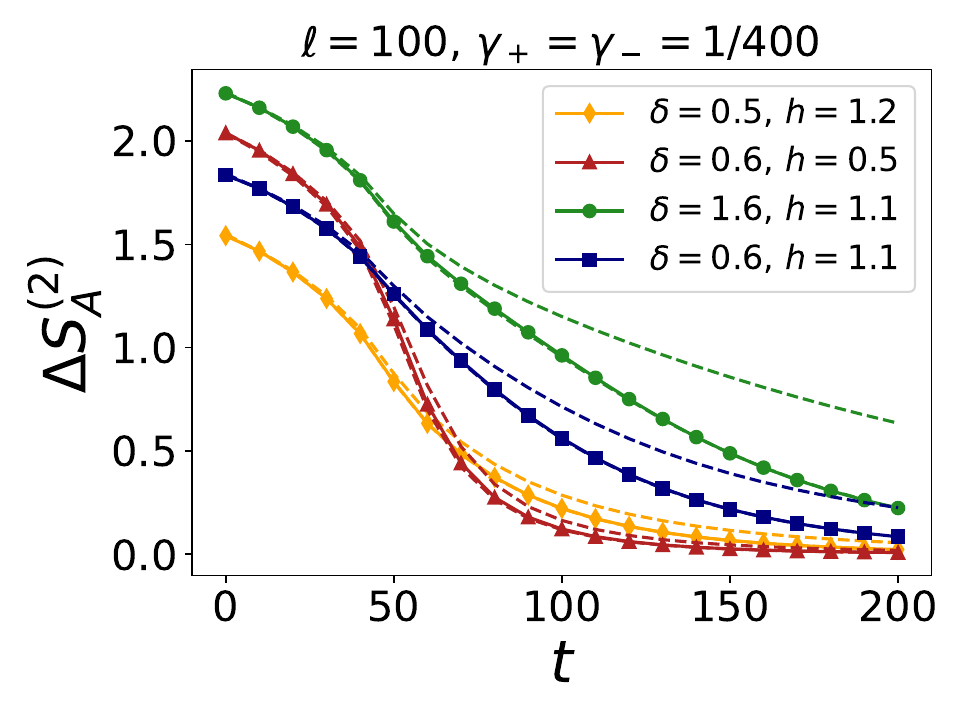}
\includegraphics[width=0.4\linewidth]{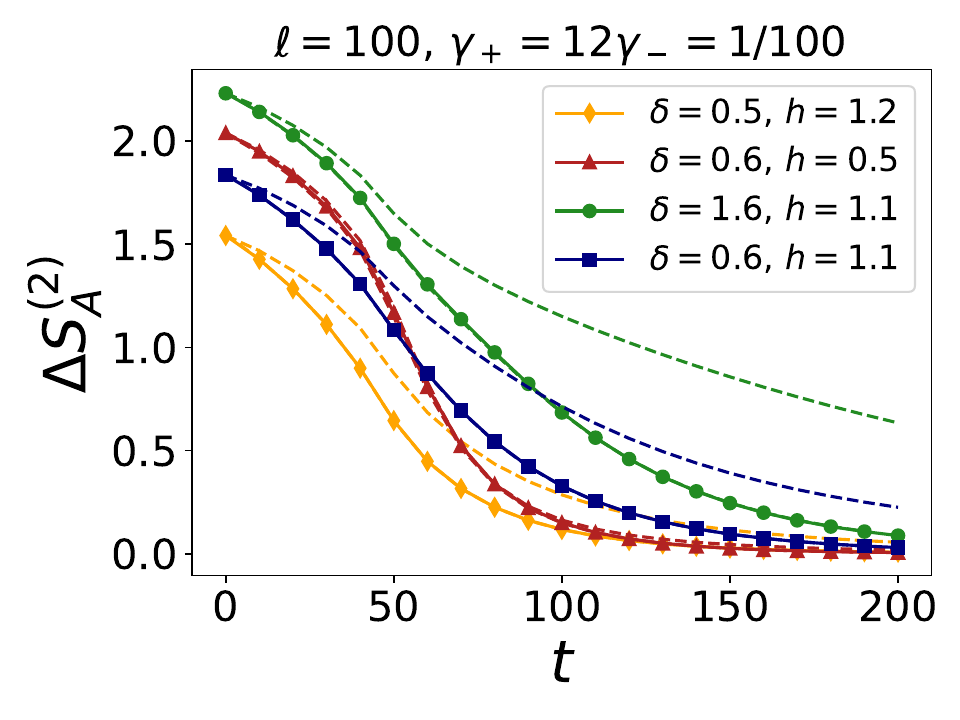}
    \caption{Time evolution of the $n=2$ R\'enyi entanglement asymmetry in a quench from the ground state of different XY spin chains in the presence of gain and loss dissipation. We consider a subsystem of length $\ell=100$ and two choices of the gain/loss rates $\gamma_\pm$. On the left, $\gamma_+=\gamma_-=1/400$; on the right, $\gamma_+=12\gamma_-=1/100$. The symbols are the exact numerical value of the entanglement asymmetry computed using Eq.~\eqref{eq:numerics} while the solid curves have been obtained with the quasiparticle prediction~\eqref{eq:charged_m}. The dashed lines denote the corresponding unitary dynamics with no dissipation, $\gamma_+=\gamma_-=0$.} \label{fig:dissfirst}
\end{figure*}

Plugging the result~\eqref{eq:charged_m} for the charged moments in Eq.~\eqref{eq:FT}, we can compute the entanglement asymmetry at any time step $t$ and any dissipation $\gamma_{\pm}\ll 1$, as we show in Fig.~\ref{fig:dissfirst}. In that figure, we study 
the time evolution of the $n=2$ R\'enyi entanglement asymmetry after
a quench from the ground state of different XY spin chains with balanced ($\gamma_+=\gamma_-$, left panel) and unbalanced gain and loss rates ($\gamma_+\neq\gamma_-$, right panel). The solid curves correspond to the quasiparticle prediction~\eqref{eq:charged_m} and the symbols are the exact numerical value calculated using the determinant formula~\eqref{eq:numerics}. We obtain an excellent agreement. To make a comparison, we also report the result due to a non-dissipative dynamics as dashed lines. The main effect of the dissipation is to diminish the entanglement asymmetry. As a consequence, the crossings of the entanglement asymmetry curves for the different initial states may be affected, undermining in some cases the QMPE. For example, the entanglement asymmetries for 
the initial couplings $(\delta, h)=(0.5, 1.2)$ and $(0.6, 0.5)$ intersect in the absence of dissipation and there is QMPE. As we can see in the plots, this intersection remains for balanced gain and loss but it disappears in the unbalanced case. On the other hand, for the pair of initial parameters $(0.6, 0.5)$ and $(0.6, 1.1)$, their entanglement asymmetries intersect both in the absence of dissipation and for balance and unbalanced gain and loss for the rates considered in Fig.~\ref{fig:dissfirst}. As we are going to discuss, a balanced gain and loss, even though suppressing the entanglement asymmetry, does not spoil the QMPE. However, in the unbalanced case, the existence of the QMPE will depend on the relative strength of the dissipation rates and the initial ground states. To see this, we have to study the long time behavior of the entanglement asymmetry.

Let us start with the balanced gain and loss dissipation. In this case, the time-evolved charged moments are given by Eqs.~\eqref{eq:charged_m}~and~\eqref{eq:larget}.  When $t\to\infty$, we observe that $x_k(\zeta)\to 0$ and, consequently, $B_{n}(\boldsymbol{\alpha},\zeta)\to 0$. Therefore, as done in detail in Ref.~\cite{rkacmb-23}, we can expand the exponential function in Eq.~\eqref{eq:charged_m} using~\eqref{eq:larget} up to the first order term in $\ell$ and, inserting it in Eq.~\eqref{eq:FT}, we compute analytically the integral in $\boldsymbol{\alpha}$. We find 
\begin{equation}\label{eq:snt}
\begin{split}
    \Delta S_A^{(n)}(t)
    & \simeq  \frac{\lambda(t)^2 \ell }{n-1}\binom{n}{2}\left[ \int_{-k^*_{\zeta}}^{k^*_{\zeta}} \frac{{\rm d}k}{2\pi}(1-2\zeta|v_k|)\sin^2\Delta_k \right.\\
  &+\left.\int_{\pi-k^*_{\zeta}}^{\pi+k^*_{\zeta}} \frac{{\rm d}k}{2\pi} (1-2\zeta|v_k|)\sin^2\Delta_k \right],
\end{split}
\end{equation}
with $k^*_{\zeta}=\arcsin{(1/(2\zeta))}$.
If we perform the change of variables $k'=k-\pi$ in the second integral, we get
\begin{equation}\label{eq:larget2}
    \Delta S_A^{(n)}(t)\simeq \frac{\lambda(t)^2 \ell n}{2} \int_{-k^*_{\zeta}}^{k^*_{\zeta}} \frac{{\rm d}k}{2\pi}(1-2\zeta|v_k|)\Upsilon_k(\delta, h).
\end{equation}
%Therefore, the condition to observe the QMPE in Eq.~\eqref{eq:mpemba_XY} does not change, not only at time $t=0$ but also for  $\zeta_I>\zeta$. 
This expression should be compared to the one in Eq.~\eqref{eq:asym_XY_tinf} for large times in the non-dissipative case. The main difference is that the balanced gain and loss dissipation introduces the factor $\lambda(t)$,  which makes the entanglement asymmetry exponentially decaying in time, instead of algebraically as in the absence of dissipation~\cite{makc-23}. The same exponential decay has been also found in Ref. \cite{cma-24}, studying the entanglement asymmetry in a setup similar to ours, but starting from the tilted N\'eel state rather than the ground state of the XY spin chain. 

With Eq.~\eqref{eq:larget2}, we can 
see how the criteria~\eqref{eq:mpemba_XY} for the occurrence of the QMPE change. Remarkably, using it in the condition~\eqref{eq:condition_mpemba_t_I}, we find that a balanced gain and loss dissipation
does not alter them, since the global factor $\lambda(t)$ does not depend on the initial state, and cancels out.

Another case where we can compute analytically $\Delta S_A^{(n)}(t)$ is when $\gamma_+\neq \gamma_-$ and $n=2$. Using Eq.~\eqref{eq:gammapm} and performing the same large time expansion as in Eq.~\eqref{eq:snt}, we get
%\begin{widetext}
%\begin{multline}
%    \Delta S_A^{(2)}(t)\simeq \lambda^2(t)(\gamma_++\gamma_-)^4\times \\\left( \int_{-k^*_{\zeta}}^{k^*_{\zeta}} \frac{{\rm d}k}{2\pi}(1-2\zeta|v_k|)\left[\frac{\sin^2\Delta_k(\gamma,h)}{(\gamma_-^2+\gamma_+^2)+(\gamma_-^2-\gamma_+^2)\cos^2\Delta_k(\gamma,h)}+\frac{\sin^2\Delta_k(\gamma,-h)}{(\gamma_-^2+\gamma_+^2)+(\gamma_-^2-\gamma_+^2)\cos^2\Delta_k (\gamma,-h)}\right]\right).
%\end{multline}
%\end{widetext}
\begin{equation}\label{eq:large_time_asymm_unbalanced}
    \Delta S_A^{(2)}(t)\simeq \frac{\ell}{4}\lambda(t)^2(\gamma_++\gamma_-)^4 \int_{-k^*_\zeta}^{k^*_\zeta} \frac{{\rm d}k}{2\pi}(1-2\zeta|v_k|)\tilde{\Upsilon}_k(\delta, h),
\end{equation}
where
\begin{equation}\label{eq:upstilde}
\begin{split}
\tilde{\Upsilon}_k(\delta,h)&=\frac{\sin^2\Delta_k(\delta,h)}{(\gamma_-^2+\gamma_+^2)+(\gamma_-^2-\gamma_+^2)\cos^2\Delta_k(\delta,h)}\\ &+
\frac{\sin^2\Delta_k(\delta,-h)}{(\gamma_-^2+\gamma_+^2)+(\gamma_-^2-\gamma_+^2)\cos^2\Delta_k(\delta,-h)}.
\end{split}
\end{equation}
When $\gamma_+=\gamma_-$, the expression above reduces to Eq.~\eqref{eq:snt} for $n=2$. Unlike for balanced gain and loss, 
the second condition in Eq.~\eqref{eq:mpemba_XY} for observing 
the QMPE gets modified if $\gamma_+\neq \gamma_-$. In fact, inserting Eq.~\eqref{eq:large_time_asymm_unbalanced} into the condition~\eqref{eq:condition_mpemba_t_I}, we obtain that the QMPE occurs if
\begin{equation}
    \label{eq:mpemba_XY2}
    \begin{cases}
        \displaystyle \int_{0}^{2\pi}\frac{{\rm d}k}{2\pi}\sin^2\Delta_k(\delta_1,h_1)>\int_{0}^{2\pi}\frac{{\rm d}k}{2\pi}\sin^2\Delta_k(\delta_2,h_2),\\
    \displaystyle   \int_{-k^*_{\zeta}}^{k^*_{\zeta}} \frac{{\rm d}k}{2\pi} \tilde{\Upsilon}_k(\delta_1,h_1)<\int_{-k^*_{\zeta}}^{k^*_{\zeta}} \frac{{\rm d}k}{2\pi} \tilde{\Upsilon}_k(\delta_2,h_2),\quad \text{for } t>t_I.
    \end{cases}
\end{equation}
Now the existence of QMPE does not depend only on the density of Cooper pairs $\sin\Delta_k$ of the initial configurations, but also on the gain and loss rates. This is also evident from the right panel of Fig. \ref{fig:dissfirst}: the crossing between the asymmetries occurs at different times with respect to the unitary evolution and, in some cases, it disappears (cf. the data for $\delta=0.5$ and $\delta=0.6$).

We observe that, if we consider a dissipative regime with only gain terms (i.e. $\gamma_-=0$), then $\tilde{\Upsilon}(\delta,h)=2/\gamma_+^2$ in Eq.~\eqref{eq:upstilde}. Since the dependence of $\Delta S_A^{(2)}(t)$ on the initial state simplifies, it means that the second condition in Eq.~\eqref{eq:mpemba_XY2} never occurs. As a consequence, we can conclude that, in the presence of only gain terms, the QMPE disappears. However, the fact that the entanglement asymmetry at large times is independent of how much the symmetry is initially broken is reminiscent of the \textit{weak} quantum Mpemba effect found in Ref.~\cite{makc-23}.

We conclude this section by commenting the nature of the stationary state of subsystem $A$. In the thermodynamic limit $N\to\infty$, the reduced-density matrix of any finite subsystem $A$ tends after the quench to a generalized Gibbs ensemble~\cite{gge,gge2, cfe-12, vr-16, fe-13}. Since $\rho_A(t)$ is Gaussian, this ensemble is univocally determined by the density of occupied modes $n(k, t)=\langle \Psi(t)|c^{\dagger}_kc_k|\Psi(t)\rangle$, reported in Eq.~\eqref{eq:occupation}, in the long time limit. In the unitary case, $n(k, t)$ is conserved by the dynamics and, therefore, is fixed by the initial state. In fact, from Eq.~\eqref{eq:occupation}, we find that, if $\gamma_{\pm}=0$, $n(k)=(1-\cos\Delta_k)/2$. This implies that the stationary state of $A$ is in that case different for each pair of initial couplings $(\delta, h)$. However, in the presence of dissipation, $n(k, t)$ varies with time and, according to Eq.~\eqref{eq:occupation}, in the long-time limit $\gamma_\pm t\to\infty$ it reads 
\begin{equation}
    n(k, t\to\infty)=\frac{1}{2}\left[1+\frac{\gamma_--\gamma_+}{\gamma_-+\gamma_+}\right],
\end{equation}
i.e. it is independent of the initial couplings $(\delta, h)$~\cite{alba2022hydrodynamics}. As a consequence, the stationary state in the presence of gain and loss terms does not depend on the initial condition of our quench. The subsystem $A$ relaxes to the same ensemble for fixed $\gamma_{\pm}$ and any $(\delta, h)$, this is the case of the quenches considered in each panel of Fig.~\ref{fig:dissfirst}. Despite the steady state is the same and does not 
depend on the initial amount of symmetry breaking, we can still observe the QMPE, as the discussion above and Fig.~\ref{fig:dissfirst} have proven.  This scenario 
is more similar to the one usually considered in the classical 
Mpemba effect, like in Ref.~\cite{kb-20}, in which the relaxation of different initial states to a common equilibrium state is studied. Therefore, we stress that, even though it is very well-known that dissipative systems as ours relax to the same unique stationary state~\cite{keize-72, nigro-19, yoshida-24}, this feature allows us to make a stronger parallelism between the classical and the quantum Mpemba effects.
%In the opposite regime, i.e. $\gamma_+= 0$, Eq. \eqref{eq:upstilde} becomes
%\begin{equation}
%    \begin{split}
%\tilde{\Upsilon}_k&(\gamma,h)=\\&\frac{\sin^2\Delta_k(\gamma,h)}{\gamma_-^2(1+\cos^2\Delta_k(\gamma,h))}+\frac{\sin^2\Delta_k(\gamma,-h)}{\gamma_-^2(1+\cos^2\Delta_k(\gamma,-h))}.
%\end{split}
%\end{equation}
%and, expanding Eq.\eqref{eq:large_time_asymm_unbalanced} around $k=0$, we find that the entanglement asymmetry reads at large time 
%\begin{equation}
%    \Delta S_A^{(2)}(t)\simeq \frac{\ell (1+h^2)\lambda^2(t)\gamma_-^2\gamma^2}{\zeta^3 96\pi (h^2-1)^2}.
%\end{equation}

\subsection{Local dephasing}\label{sub:dephasing}

So far, we have focused our attention on a dissipative evolution induced by gain and loss terms. We now want to study what is the effect of another source of dissipation, which is the local dephasing, modeled by the jump operator $L_{j,z}=\sqrt{\gamma_d}a^{\dagger}_ja_j$. In this case, the dynamics does not respect the Gaussianity of the initial state. However, for small dephasing rates $\gamma_d\ll 1$,
we can take as an approximation of $\rho(t)$ the Gaussian state determined by the two-point correlation matrix~\eqref{eq:def_corr}~\cite{alba2021spreading}. Under this assumption, we can calculate the charged moments and, therefore, the entanglement asymmetry with Eq.~\eqref{eq:numerics}. To this end, in Appendix \ref{app:sub2}, we derive the equations of motion for the two-point functions \( G_{mn}=\langle a^{\dagger}_ma_n \rangle \) and \( F_{mn}=\langle a^{\dagger}_ma^{\dagger}_n \rangle \) in systems of free fermions with dephasing, building upon the findings in Ref.~\cite{alba2021spreading, Eislerdephasing2011,Albadephasing2023, prosen2008}. We show that, since both correlations satisfy linear first order differential equations (see Eqs.~\eqref{eq:evolg} and~\eqref{eq:evolf}), they can be treated efficiently as a matrix eigenvalue problem. 
By numerically solving them, we can build the correlation matrix defined in Eq.~\eqref{eq:def_corr}, from which we compute the charged moments using Eq.~\eqref{eq:numerics}. Plugging this result in Eq.~\eqref{eq:FT}, we obtain the main result of this section, which is the time evolution of the entanglement asymmetry in the presence of local dephasing. We remark that,  although the Gaussianity of the state is not 
respected by the dynamics at all times, we expect that the deviations from Gaussian behavior are small as $\gamma_d\ll 1$, which is the regime also considered in~\cite{alba2021spreading} to analyze the effects of local dephasing in the entanglement entropy in a similar setting. 

\begin{figure*}
\centering
\includegraphics[width=0.32\linewidth]{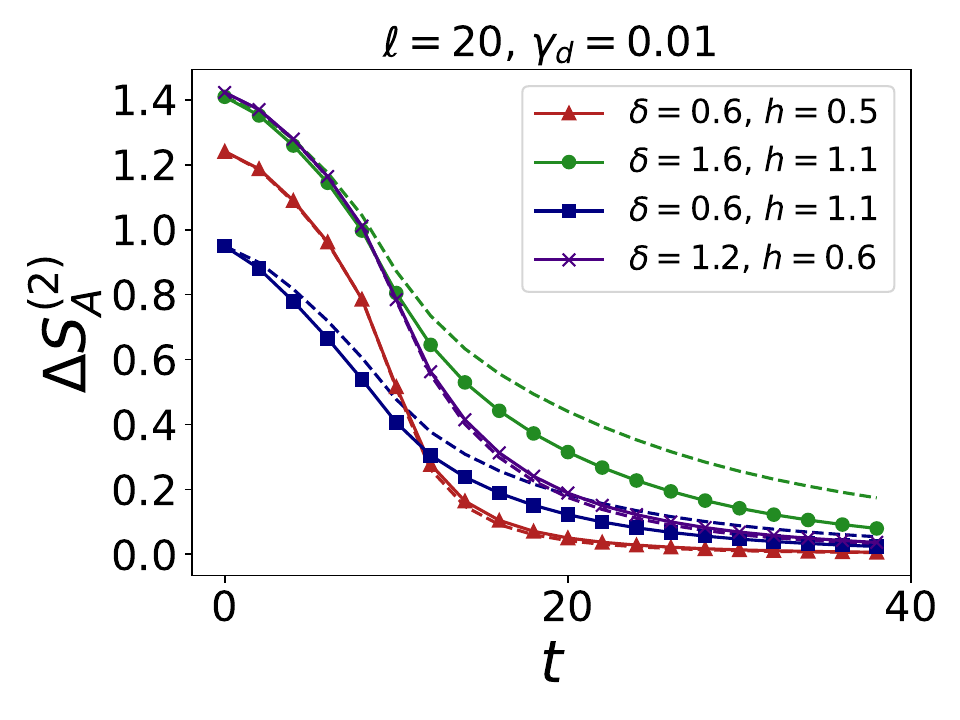}\includegraphics[width=0.32\linewidth]{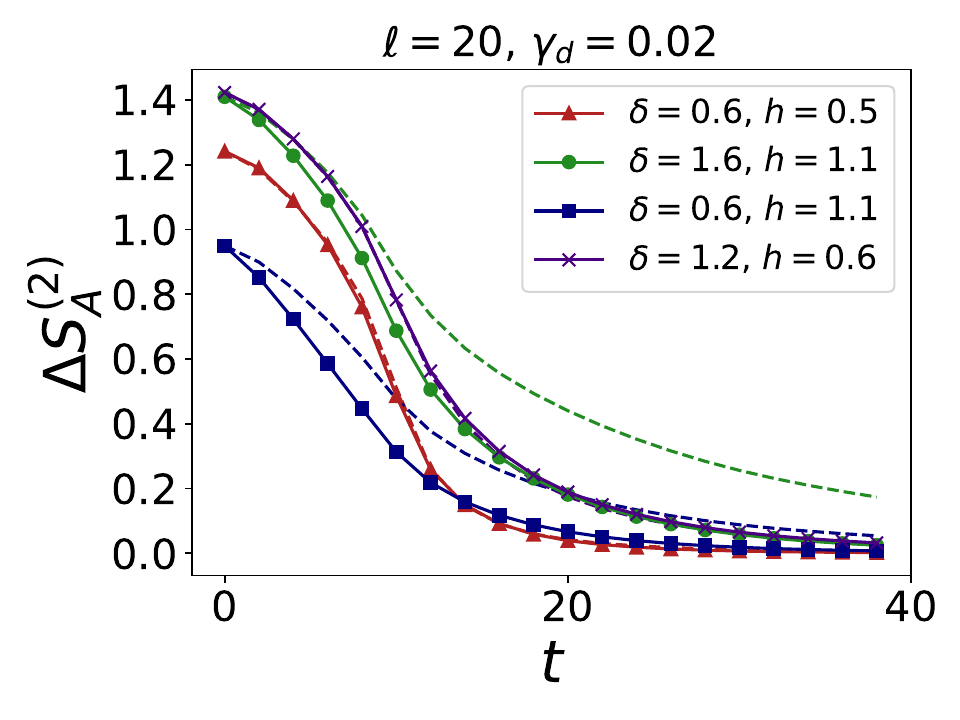}\includegraphics[width=0.32\linewidth]{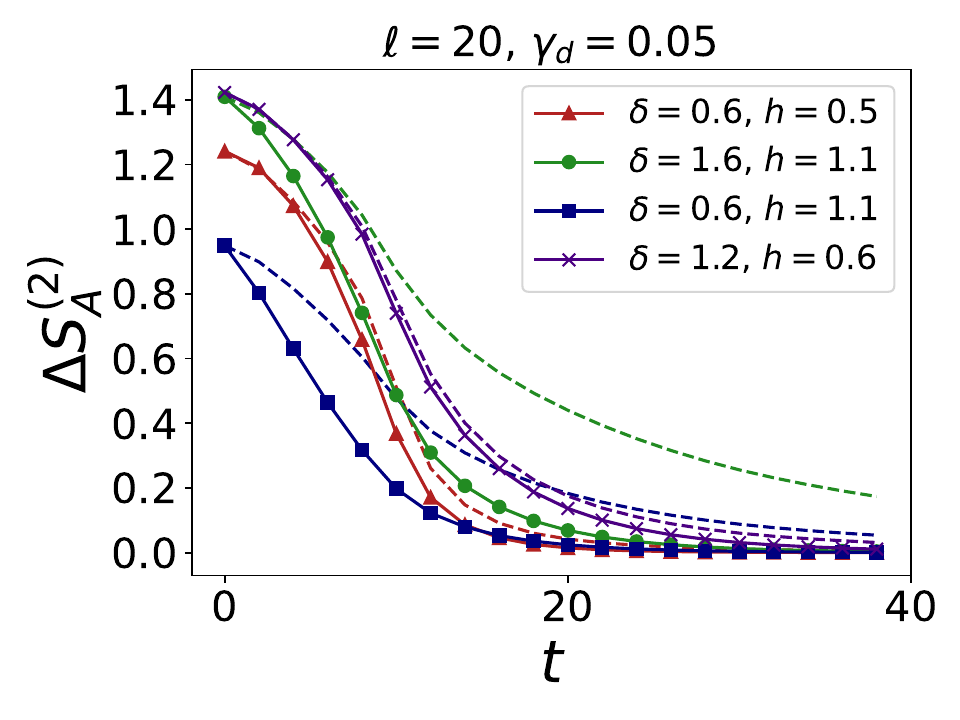}
\caption{Time evolution of the $n=2$ R\'enyi entanglement asymmetry
after a quench from the ground state of different XY spin chains to the XX spin chain in the presence of local dephasing. In each panel, we take a different dephasing rate $\gamma_d= 0.01,0.02,0.05$ and the same initial states.  In all the cases, the subsystem $A$ is an interval of length $\ell=20$ in a chain of total size $N=10\ell$.  The symbols have been obtained numerically calculating the two-point correlation functions at each time step with Eqs.~\eqref{eq:evolg}-\eqref{eq:evolf} and then applying Eq.~\eqref{eq:numerics}. The dashed lines correspond to the unitary evolution ($\gamma_d=0$). In this case, the solid lines only join the numerical symbols as a guide for the eyes.}
\label{fig:ff_dephasing}
\end{figure*}

We show the dynamics of the $n=2$ R\'enyi entanglement asymmetry in Fig.~\ref{fig:ff_dephasing}, where each panel corresponds to a different value of the dephasing rate $\gamma_d=0.01,0.02,0.05$. In all the panels, we consider a subsystem $A$ of length $\ell=20$ in a system of size $N=10\ell$ and four initial conditions, corresponding to the ground state of the XY Hamiltonian in Eq.~\eqref{eq:Ham_XY} with different parameters $\delta$ and $h$. For comparison, we report the unitary dynamics without dissipation ($\gamma_d=0$) as dashed lines. As in the case of gain an loss, see Fig.~\ref{fig:dissfirst}, the local dephasing induces a decrease in the entanglement asymmetry. This effect is more marked for the initial states with $h>1$ than for those with $h<1$, for which we need
larger dephasing rates to see a clear difference with respect to the
unitary dynamics (dashed lines). As a result, the time at which the 
entanglement asymmetries intersect may be shifted towards larger 
times.  
A similar suppression of the entanglement 
asymmetry with a global dephasing has been observed in the 
experimental work~\cite{joshi-24}, where an ion trap simulates the dynamics of a XX spin chain with long-range couplings that is initially prepared in a tilted ferromagnetic state. These configurations are precisely the ground state of the XY Hamiltonian~\eqref{eq:Ham_XY} along the curve $\delta^2+h^2=1$~\cite{ktm-82, ms-85}. 
In the experimental setting, the crossing of the curves describing the entanglement asymmetry remains almost unaffected. 

%and also the crossing time at which the QMPE occurs is shifted towards larger time. These features can be also observed in Fig.~\ref{fig:dissfirst}, implying that even gain and loss terms favour a decrease of $\Delta S_A^{(2)}$ and a shift of the crossing time at larger time. However, for $h<1$, the red lines of the three panels suggest that we need larger values of the dephasing in order to see a clear difference with respect to the unitary dynamics (dashed lines).
%This conclusion is similar to what has been observed in Ref. \cite{joshi-24}, which has experimentally studied the entanglement asymmetry in the presence of the global dephasing: by choosing an initial state satisfying $\gamma^2+h^2=1$ (i.e. $h<1$), they found that the crossing of the curves describing the entanglement asymmetry remains almost unaffected.  
%To summarize, the effect of dephasing (both global and local) is similar and even small values of $\gamma_d$ play an important role especially for $h>1$. %
The physical origin behind this difference might be due to the fact that, for $h<1$, the initial state is in an “ordered phase”, with a net magnetization along the $z$-axis. Therefore, this contrasts the effect of the dephasing, which tends to rotate the spins around the $z$-axis during the time evolution. On the other hand, when the initial state is in the disordered phase, i.e. $h>1$, there is no net magnetization along $z$ and the spins are less constrained in the dynamics, so smaller values of the dephasing are enough to observe a major effect on the entanglement asymmetry. 
%The different result about the shifting of the crossing time we observe is due to the effect of a local rather than a global dissipation. 

\section{Global quench from a finite temperature state}\label{sec:temperature}

In this section, we consider the evolution of the entanglement asymmetry in a quantum quench from the XY spin chain~\eqref{eq:Ham_XY} at a finite temperature $1/\beta$ to the XX spin chain. We keep the whole system isolated from the environment. In this case, the system is initially described by the Gibbs ensemble $\rho_\beta=e^{-\beta H_{\rm XY}}/Z$, where $Z={\rm Tr}(e^{-\beta H_{\rm XY}})$. After the quench, this state evolves unitarily as $\rho_\beta(t)=e^{-it H_{\rm XX}} \rho_\beta e^{itH_{\rm XX}}$. 

Since both the initial state $\rho_\beta$ and the post-quench Hamiltonian are Gaussian, the dynamics of the system is fully encoded in the spatial two-point fermionic correlations~\eqref{eq:def_corr}. For the thermal state, the two-point correlation matrix in the thermodynamic limit $N\to\infty$ is also a block Toeplitz matrix of the form of Eq.~\eqref{eq:Gammat0} where $\mathcal{G}_0(k, t)$ is replaced by the $2\times 2$ symbol
\begin{equation}\label{eq:symbol_corr_temp}
\mathcal{G}_\beta(k, t)=C_{\beta, k}
\left(\cos\Delta_k \sigma^z + \sin \Delta_k \sigma^y e^{2it \epsilon_k \sigma^y}\right),
\end{equation}
and $C_{\beta, k}=\tanh(\beta \epsilon_k^{\rm XY}/2)$ and $\epsilon_k^{\rm XY}$ is the dispersion relation of the Hamiltonian of the XY spin chain~\eqref{eq:Ham_XY}, $\epsilon_k^{\rm XY}=\sqrt{(h-\cos k)^2+\delta^2\sin^2k}$.
\bigskip
\subsection{Entanglement asymmetry at finite temperature}

Let us first analyze the entanglement asymmetry in the Gibbs ensemble 
$\rho_\beta$. To obtain the asymptotic behavior of the charged moments $Z_n(\boldsymbol{\alpha}, \beta)$, we can apply its expression (Eq.~\eqref{eq:numerics}) in terms of the two-point correlation matrix together with Eq.~\eqref{eq:conj_1}, specialized to the finite temperature symbol (Eq.~\eqref{eq:symbol_corr_temp}) at $t=0$. We find that
\begin{equation}\label{eq:charged_mom_gibbs}
Z_n(\boldsymbol{\alpha}, \beta)= Z_n(\boldsymbol{0}, \beta)
e^{B_n(\boldsymbol{\alpha}, \beta)\ell},
\end{equation}
where the coefficient $B_n(\boldsymbol{\alpha}, \beta)$ reads for $n=2$
\begin{equation}\label{eq:B_2_t0}
B_2(\alpha, \beta)=\int_{0}^{2\pi}\frac{{\rm d} k}{4\pi}\log\left[1-\frac{4\sin^2\alpha \sin^2\Delta_k}{(C_{\beta, k}^{-1}+C_{\beta, k})^2}\right]
\end{equation}
and, for $n=3$,
\begin{widetext}
\begin{equation}\label{eq:B_3_t0}
B_3(\boldsymbol{\alpha}, \beta)=\int_{0}^{2\pi} \frac{{\rm d}k}{4\pi}
\log\left[1 - \frac{4 C_{\beta, k}^2(1 + C_{\beta, k}^2) \sin^2\Delta_k \sum_{j=1}^3\sin^2\alpha_{j, j+1} - 
     16 i \cos\Delta_k\sin^2\Delta_k C_{\beta, k}^3 \prod_{j=1}^3\sin \alpha_{j, j+1}}{(1 + 
      3 C_{\beta, k}^2)^2}\right].
\end{equation}
\end{widetext}
When we take $\beta\to\infty$ in the previous equations, we recover the result for the ground state, Eq.~\eqref{eq:B_n_unitary} at $t=0$. The expression of $B_n(\boldsymbol{\alpha}, \beta)$ is more and more involved as we increase $n$ and we have not been able to get a closed analytic form for integer $n$ as in the zero temperature limit. 
%In fact, when $\beta\to\infty$, one recovers the result for the ground state, Eq.~\eqref{eq:B_n_unitary} at $t=0$. \sara{This equation shows that $B_n(\boldsymbol{\alpha}, \beta)$ decomposes in the sum over the replicas $n$, so it means that in the pure state limit $\beta \to \infty$, we can recover this known result.}
%which decomposes in the sum over the replicas $n$ that we consider.  
We can see that, while at zero temperature the charged moments factorize in the replica space, i.e. the exponent $B_n(\boldsymbol{\alpha},\beta\to\infty)$ decomposes as in Eq.~\eqref{eq:fact_rep_space}, at finite $\beta$, $B_n(\boldsymbol{\alpha},\beta)$ does not generically satisfy such decomposition. This does not come as a surprise because also in the presence of gain and loss, we have shown that Eq.~\eqref{eq:B_n_dissipation} does not admit a decomposition in terms of each replica.
% as also happens for Eq.~\eqref{eq:init_time}. 

To obtain the R\'enyi entanglement asymmetry from the charged moments in Eq.~\eqref{eq:charged_mom_gibbs}, we plug them into Eq.~\eqref{eq:FT}. 
The $n$-fold integral can be exactly computed for large subsystem size $\ell$ doing a saddle point approximation. The calculation follows the same steps as in the ground state case described in detail in Ref.~\cite{makc-23}. 

The final result~is
\begin{equation}\label{eq:saddle_point_ent_asymm_gibbs}
\Delta S_A^{(n)}(\delta, h,\beta)=\frac{1}{2}\log \ell +\frac{1}{2}\log\frac{\pi  g_\beta^{(n)}(\delta, h)n^{\frac{1}{1-n}}}{4}+O(\ell^{-1}).
\end{equation}

 As in the ground state, see Eq.~\eqref{eq:asym_XY_t0}, the entanglement asymmetry grows logarithmically with the subsystem size $\ell$ and the same coefficient $1/2$, related to the dimension of the $U(1)$ group. The temperature modifies the $\ell$-independent function $g_\beta^{(n)}(\delta, h)$, cf. Eq.~\eqref{eq:g_gs},  which now depends on the R\'enyi index $n$ and the temperature in a non-trivial way, as a consequence of the non factorization of the charged moments (Eq.~\eqref{eq:charged_mom_gibbs}) in the replica space. For $n=2$, we obtain
\begin{equation}
g_\beta^{(2)}(\delta, h)=\int_{0}^{2\pi}\frac{{\rm d}k}{2\pi}\tanh^2(\beta\epsilon_k^{\rm XY})\sin^2\Delta_k(\delta, h),
\end{equation}
while, for $n=3$, 
\begin{equation}
g_\beta^{(3)}(\delta, h)=\int_{0}^{2\pi}\frac{{\rm d}k}{2\pi}
\frac{8C_{\beta, k}^2(1+C_{\beta, k}^2)\sin^2\Delta_k(\delta, h)}{(1+3C_{\beta, k}^2)^2}.
\end{equation}
Notice that, in the limit $\beta\to \infty$, we recover the prediction for the ground state in Eq.~\eqref{eq:g_gs}. In the infinite temperature limit, $\beta=0$, the Gibbs ensemble reduces to the (normalized) identity, $\rho_{\beta=0}=2^{-N} I$, which commutes with the charge $Q$, and, therefore, $\Delta S_A^{(n)}(\delta, h, \beta=0)=0$. However, if we take $\beta=0$ in the saddle point approximation (Eq.~\eqref{eq:saddle_point_ent_asymm_gibbs}), the expression diverges. The reason is that the limits $\ell\to \infty$ and $\beta\to 0$ do not commute. To obtain the behavior of the R\'enyi entanglement asymmetry at large temperatures, we can
expand the expression~\eqref{eq:charged_mom_gibbs} for the charged moments around $\beta=0$ keeping $\ell$ finite. If we truncate the expansion at order $O(\beta^2)$, then all the integrals can be straightforwardly calculated. We eventually find that, in the limit $\beta\to 0$, the R\'enyi entanglement asymmetry behaves for $n=2$ as
\begin{equation}\label{eq:large_temp_gibbs_n_2}
\Delta S_A^{(2)}(\delta, h,\beta)\simeq \frac{\ell \delta^2 \beta^2}{8}
\end{equation}
and for $n=3$
\begin{equation}\label{eq:large_temp_gibbs_n_3}
\Delta S_A^{(3)}(\delta, h, \beta)\simeq \frac{3\ell \delta^2\beta^2}{16}.
\end{equation}

\begin{figure}[t]
\includegraphics[width=\linewidth]{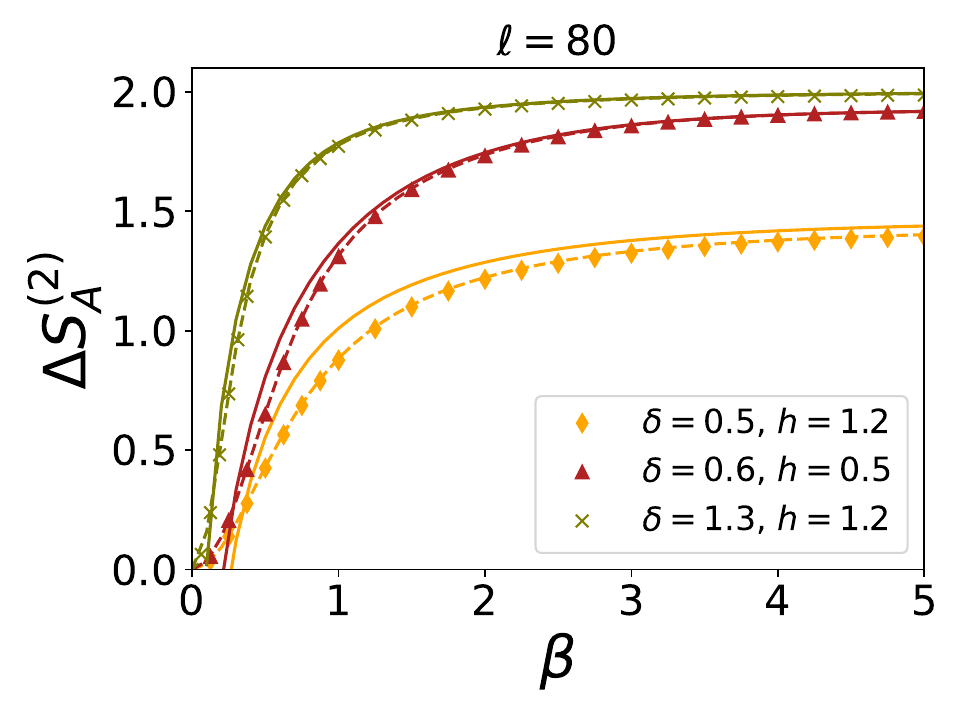}   
    \caption{$n=2$ R\'enyi entanglement asymmetry in the Gibbs ensemble of the XY spin chain~\eqref{eq:Ham_XY} in the thermodynamic limit $N\to\infty$ as a function of the inverse temperature $\beta$ for different values of the couplings $\delta$ and $h$. In all the cases, we take as subsystem $A$ an interval of length $\ell=80$. The symbols are the exact value of the entanglement asymmetry, calculated numerically using Eq.~\eqref{eq:numerics}. The solid curves represent the saddle point approximation in Eq.~\eqref{eq:saddle_point_ent_asymm_gibbs} and the dashed curves have been obtained using the analytic prediction for the charged moments~\eqref{eq:charged_mom_gibbs}, without applying the saddle point approximation.}\label{fig:asymm_n_2_temp}
\end{figure}

In Fig.~\ref{fig:asymm_n_2_temp}, we represent the $n=2$ R\'enyi entanglement asymmetry 
as a function of the inverse temperature $\beta$ for different values 
of the couplings $\delta$ and $h$ for a subsystem of length $\ell=80$. The symbols are the exact entanglement asymmetry calculated using Eq.~\eqref{eq:numerics}, the solid curves correspond to the asymptotic
expression of Eq.~\eqref{eq:saddle_point_ent_asymm_gibbs}, and the dashed curves have been obtained calculating exactly the Fourier transformation (Eq.~\eqref{eq:FT}) of the charged moments in Eq.~\eqref{eq:charged_mom_gibbs} without taking the saddle point approximation. We observe that the entanglement asymmetry is a monotonic decreasing function
of the temperature that vanishes, as expected, in the limit $\beta\to0$, at which the symmetry is recovered.
\\

\subsection{Entanglement asymmetry after the quench}

We now move on to the evolution of the R\'enyi entanglement 
asymmetry after a quench to the XX spin chain from the Gibbs ensemble 
$\rho_\beta$ of the XY Hamiltonian. When $N\to \infty$, the $U(1)$ particle number symmetry initially broken 
by $\rho_\beta$ is locally restored at $t\to\infty$. In fact, in that limit, the time dependent term $e^{i 2t \epsilon_k \sigma^y}$ in the symbol~\eqref{eq:symbol_corr_temp} of the time-evolved two-point correlation matrix averages to zero and
\begin{equation}
\mathcal{G}_\beta(k,t\to \infty)= C_{\beta, k} \cos\Delta_k \sigma^z.
\end{equation}
If we calculate the charged moments using this symbol and applying Eqs.~\eqref{eq:numerics}~and~\eqref{eq:conj_1}, we find that $Z_n(\boldsymbol{\alpha},\beta, t\to\infty)=Z_n(\boldsymbol{0}, \beta, t\to\infty)$, which implies that the R\'enyi entanglement asymmetry vanishes at $t\to \infty$.

To derive the exact full time evolution of $Z_n(\boldsymbol{\alpha}, \beta, t)$ after the quench, we can proceed as explained in Appendix~\ref{app:sub1} 
for the case of gain and loss dissipation and resort to the quasiparticle picture. According to it, the time evolved charged moments are given by Eq.~\eqref{eq:charged_m} upon replacing the symbol $\mathcal{G}(k, t\to 0)$ in the matrix $\mathcal{M}_{\boldsymbol{\alpha}}^{(n)}(k,t\to0)$ by the finite temperature symbol $\mathcal{G}_\beta(k,t=0)$ at time zero. Therefore, in the hydrodynamic limit $t,\ell\to \infty$ with $t/\ell$ fixed,
\begin{equation}\label{eq:charged_moments_gibbs_time}
Z_n(\boldsymbol{\alpha}, \beta, t)=Z_n(\boldsymbol{0}, \beta, t) e^{B_n(\boldsymbol{\alpha}, \beta, \zeta)\ell}.
\end{equation}
\bigskip
The explicit expression of $B_n(\boldsymbol{\alpha},\beta, \zeta)$ for $n=2$ is

\begin{equation}\label{eq:B_2_t}
B_2(\alpha, \beta, \zeta)=\int_{0}^{2\pi}\frac{{\rm d} k}{4\pi} x_k(\zeta)
\log\left[1-\frac{4\sin^2\Delta_k \sin^2\alpha}{(C_{\beta, k}^{-1}+C_{\beta, k})^2}\right],
\end{equation}
and for $n=3$
\begin{widetext}
\begin{equation}\label{eq:B_3_t}
B_3(\boldsymbol{\alpha}, \beta, \zeta)=\int_{0}^{2\pi} \frac{{\rm d}k}{4\pi}
x_k(\zeta)\log\left[1 - \frac{4 C_{\beta, k}^2(1 + C_{\beta, k}^2) \sin^2\Delta_k \sum_{j=1}^3\sin^2\alpha_{j, j+1} - 
     16 i \cos\Delta_k\sin^2\Delta_k C_{\beta, k}^3 \prod_{j=1}^3\sin\alpha_{j, j+1}}{(1 + 
      3 C_{\beta, k}^2)^2}\right].
\end{equation}
\end{widetext}
As we have already explained, the term $x_k(\zeta)$ in $B_n(\boldsymbol{\alpha}, \beta, t)$ accounts for the number of entangled pairs of quasiparticles created at the quench and with opposite velocity $\pm v_k$ that are inside the subsystem $A$ at time
$t$. The other term in the integrand is the contribution of these excitations to the ratio $Z_n(\boldsymbol{\alpha}, \beta, t)/Z_n(\boldsymbol{0},\beta, t)$ which, since in this case the evolution is unitary, is determined by their value at $t=0$, compare  Eqs.~\eqref{eq:B_2_t} and~\eqref{eq:B_3_t} with Eqs.~\eqref{eq:B_2_t0} and~\eqref{eq:B_3_t0}.  

\begin{figure*}[t]
    \includegraphics[width=0.32\linewidth]{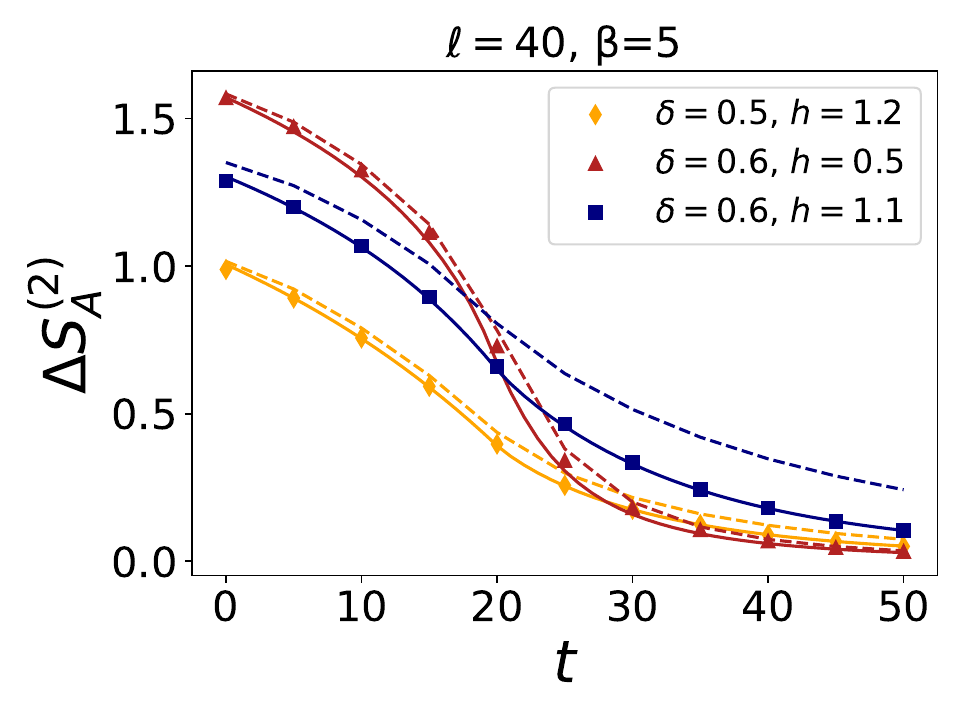}
    \includegraphics[width=0.32\linewidth]{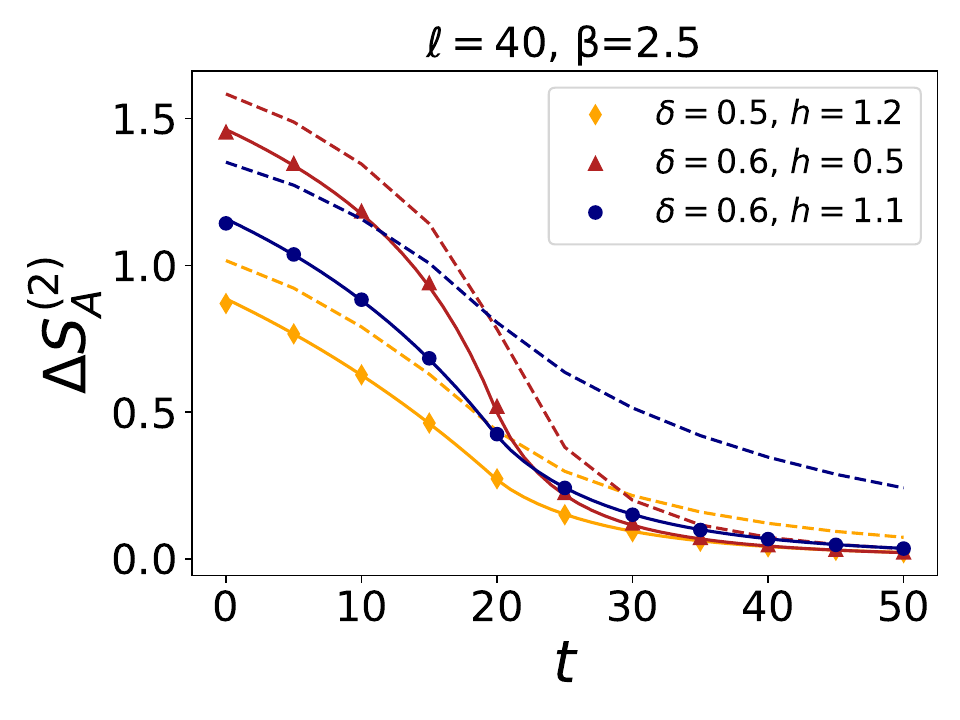}
    \includegraphics[width=0.32\linewidth]{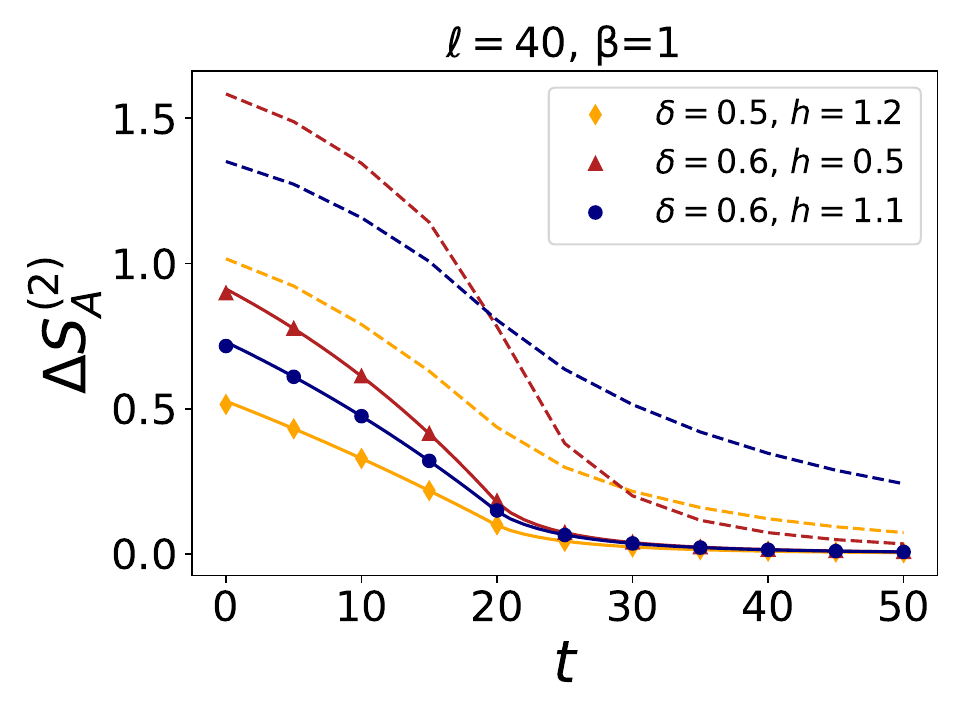}
    \caption{Time evolution of the $n=2$ R\'enyi entanglement asymmetry after a quench to the XX spin chain from a XY spin chain at inverse temperature $\beta=5$ (right panel), $2.5$ (middle panel) and $1$ (left panel). In all the panels, we consider the same sets of values for the parameters $h$ and $\delta$ of the initial XY spin chain and we take a subsystem of size $\ell=40$. The symbols are the exact entanglement asymmetry, obtained numerically using the determinant formula~\eqref{eq:numerics}, and the solid curves correspond to the quasiparticle prediction~\eqref{eq:charged_moments_gibbs_time} for the charged moments. The dashed lines correspond to the exact time evolution from the chain at zero temperature ($\beta \rightarrow \infty$).}
    \label{fig:asymm_quench_gibbs_n_2}
\end{figure*}

\begin{figure*}[t]
    \includegraphics[width=0.32\linewidth]{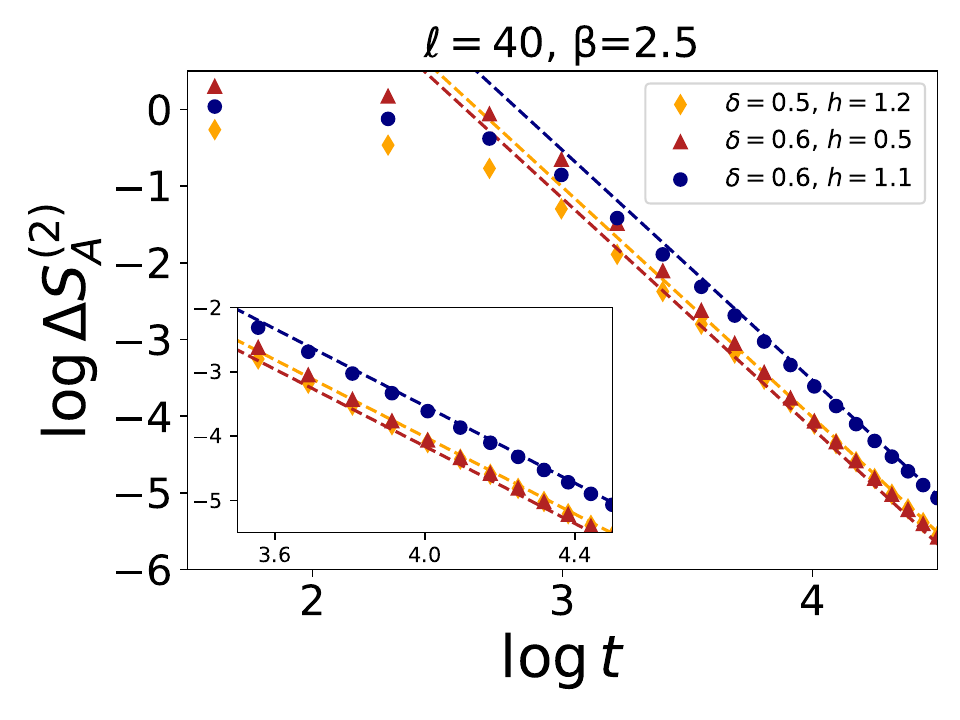}
    \includegraphics[width=0.32\linewidth]{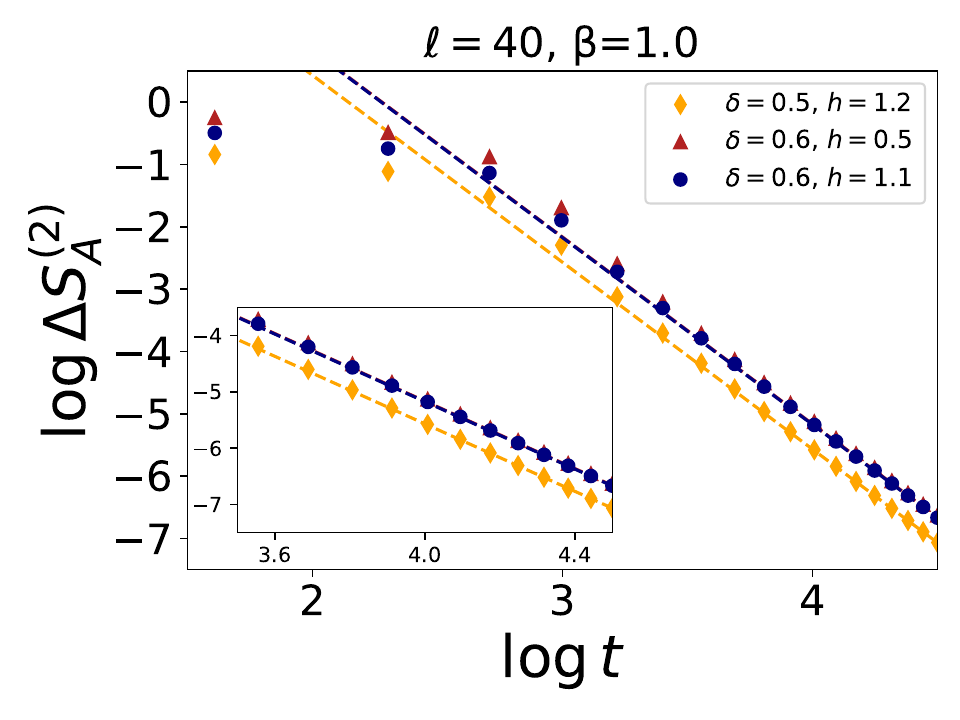}
    \includegraphics[width=0.30\linewidth]{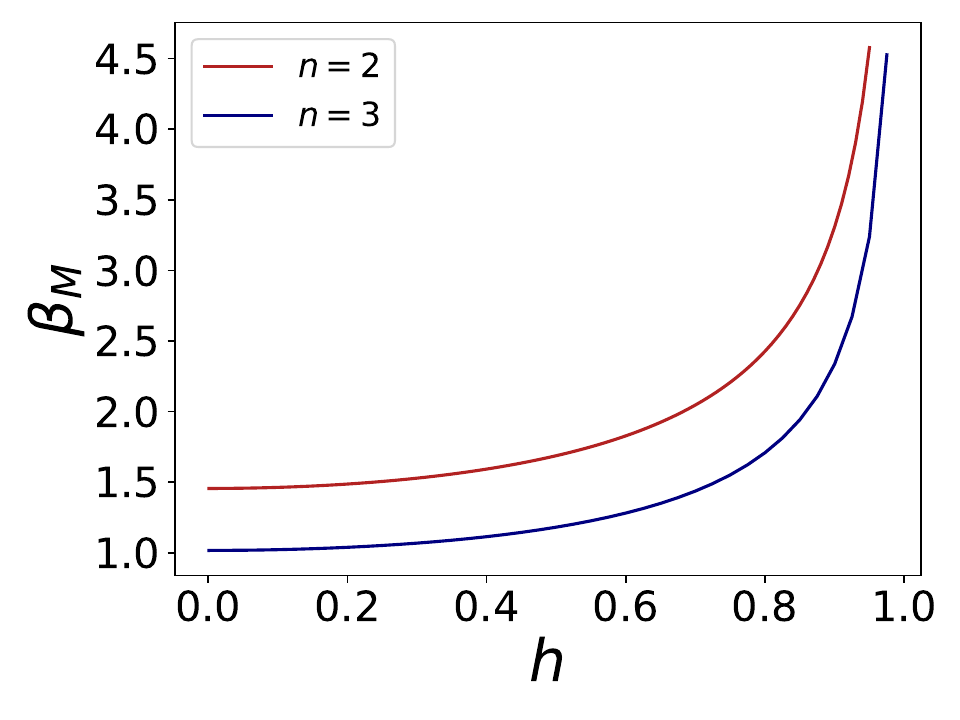}
    \caption{The left and middle panels are the same plot of middle and right panels of Fig.~\ref{fig:asymm_quench_gibbs_n_2}, but taking the logarithm in both axes to check that the entanglement asymmetry decays at large times as predicted by Eq.~\eqref{eq:asymm_temp_asymp_n_2}. 
    The symbols are the exact value of the entanglement asymmetry while the solid lines correspond to Eq.~\eqref{eq:asymm_temp_asymp_n_2}. The inset is a zoom of the large time region.
    In the righ panel, temperature of the initial XY spin chain from which the QMPE ceases to exist. We consider the quenches from two different chains with couplings $h_1=0$, $\delta_1=1$ and $h_2=h$, $\delta_2=\sqrt{1-h^2}$. The curves represent the solution
     $\beta_{\rm M}$  as a function of $h$ of the equation $\Delta S_A^{(n)}(h_1=0, \delta_1=1, \beta_{\rm M}, t)=\Delta S_A^{(n)}(h, \sqrt{1-h^2}, \beta_{\rm M}, t)$ at large times using Eq.~\eqref{eq:asymm_temp_asymp_n_2} for $n=2$ (red curve) and and Eq.~\eqref{eq:asymm_temp_asymp_n_3} $n=3$ (blue curve).}
    \label{fig:beta_M}
\end{figure*}

Inserting Eq.~\eqref{eq:charged_moments_gibbs_time} in Eq.~\eqref{eq:FT}, we obtain the time evolution of the R\'enyi entanglement asymmetry after the quench. In Fig.~\ref{fig:asymm_quench_gibbs_n_2},  we represent the R\' enyi entanglement asymmetry for $n=2$ as a function of time for different values of the temperature in the initial chain, $\beta=5$, $2.5$ and $1$, taking three different sets of couplings $(\delta, h)$ of the XY Hamiltonian. The solid curves have been obtained with the quasiparticle prediction for the charged moments as in Eq.~\eqref{eq:charged_moments_gibbs_time} and the symbols are the exact value of the entanglement 
asymmetry calculated numerically using the determinant formula in Eq.~\eqref{eq:numerics}. For comparison, we also include as dashed lines the time evolution from the corresponding ground states, that is, the limit $\beta\to\infty$.
The pairs of couplings $(0.5, 0.6)$ and $(0.5, 1.2)$ or $(0.6, 1.1)$ satisfy the conditions of Eq.~\eqref{eq:mpemba_XY} and, in a quench from the corresponding ground states to the XX spin chain, their R\'enyi entanglement asymmetries intersect at a finite time, indicating the occurrence of the QMPE. On the other hand, the couplings $(0.5, 1.2)$ and $(0.6, 1.1)$ violate Eqs.~\eqref{eq:mpemba_XY} and their ground state entanglement asymmetries do not intersect; therefore, they do not show QMPE. When quenching from a chain at finite temperature, we observe that the QMPE still occurs at low enough temperatures for those pairs of couplings for which there is QMPE at zero temperature (left panel of Fig.~\ref{fig:asymm_quench_gibbs_n_2}). However, as we take a larger temperature, the difference between their entanglement asymmetries at late times progressively decreases. There is a specific temperature from which the entanglement asymmetries do not cross anymore and the QMPE disappears. This can be clearer observed in the left and middle panels of Fig.~\ref{fig:beta_M}, in which have replotted the middle and right panels of Fig.~\ref{fig:asymm_quench_gibbs_n_2}, but taking the logarithm of both axes. At $\beta=1$, the crossing between the asymmetries for the pair of couplings (0.5, 1.2) and (0.6, 0.5) has disappeared. On the other hand, for that initial temperature, the asymmetries for the pairs (0.6, 0.5) and (0.6, 1.1) do not intersect, but they overlap at large $t$. As we will see, when $\beta\ll 1$, the large time behavior of the asymmetry only depends on the initial value of the anisotropy parameter $\delta$ of the XY spin chain, but it is independent of the transverse magnetic field $h$.

To clarify this phenomenology and establish the conditions
for the occurrence of the QMPE in the quench from the Gibbs ensemble, let us analyze the large time behavior of the R\'enyi entanglement asymmetry as in the previous sections. Again, in the coefficient $B_n(\boldsymbol{\alpha}, \beta, \zeta)$ of the charged moments, the term $x_k(\zeta)$ vanishes for $2\zeta|v_k|>1$ and, therefore, only the modes with the slowest group velocity, those around $k=0$ and $\pi$, contribute when we approach the equilibrium. We can then expand the logarithm function in Eqs.~\eqref{eq:B_2_t} and~\eqref{eq:B_3_t} around $k=0$ and $\pi$. Since $B_n(\boldsymbol{\alpha}, \beta, \zeta)\to 0$ at $\zeta\to\infty$, we can further take the Taylor expansion of the exponential function in 
Eq.~\eqref{eq:charged_moments_gibbs_time} and, truncating it at the first order term, calculate exactly the integral in $\boldsymbol{\alpha}$, as we have done in Eq.~\eqref{eq:snt} for the case of gain and loss dissipation.  We eventually find
\begin{equation}\label{eq:large_time_asymm_gibbs}
\Delta S_A^{(n)}(t)\simeq \frac{n \ell}{n-1} \int_{-k^*_{\zeta}}^{k^*_{\zeta}}\frac{{\rm d}k}{4\pi} (1-2\zeta|v_k|)\Upsilon_{\beta, k}^{(n)}(\delta, h),
\end{equation}
where the function $\Upsilon_{\beta, k}^{(n)}(\delta, h)$ for $n=2$ is
\begin{multline}\label{eq:Upsilon_beta_2}
\Upsilon_{\beta,k}^{(2)}(\delta, h)=\frac{ \sin^2\Delta_k(\delta, h)}{(C_{\beta, k}(\delta, h)^{-1}+C_{\beta, k}(\delta, h))^2}\\+\frac{\sin^2\Delta_k(\delta, -h)}{(C_{\beta, k}(\delta, -h)^{-1}+C_{\beta, k}(\delta, -h))^2},
\end{multline}
and for $n=3$
\begin{multline}\label{eq:Upsilon_beta_3}
\Upsilon_{\beta, k}^{(3)}(\delta, h)=\frac{C_{\beta, k}(\delta, h)^2(1+C_{\beta, k}(\delta, h)^2)\sin^2\Delta_k(\delta, h)}{(1+3 C_{\beta, k}(\delta, h)^2)^2}\\
+\frac{C_{\beta, k}(\delta, -h)^2(1+C_{\beta, k}(\delta, -h)^2)\sin^2\Delta_k(\delta, -h)}{(1+3 C_{\beta, k}(\delta, -h)^2)^2}.
\end{multline}
The function $\Upsilon_{\beta,k}^{(n)}(\delta, h)$ represents the contribution of the slowest modes to the entanglement asymmetry at 
large times. As clear from Eqs.~\eqref{eq:Upsilon_beta_2} and~\eqref{eq:Upsilon_beta_3}, its explicit expression
depends on the R\'enyi index $n$ in a very non-trivial way that we have not been able to disentangle. In the zero temperature limit $\beta\to \infty$, $\Upsilon_{\beta, k}^{(n)}(\delta, h)$  tends to $\Upsilon_k(\delta, h)$, which was defined in Eq.~\eqref{eq:Upsilon_gs} and is independent of $n$. 

Let us now consider two different initial Gibbs ensembles $\rho_{\beta}(\delta_1, h_1)$ and $\rho_{\beta}(\delta_2, h_2)$ for which $\Delta S_A^{(n)}(\delta_1, h_1, \beta)>\Delta S_A^{(n)}(\delta_2, h_2,\beta)$. If we plug the asymptotic expression of their entanglement asymmetries at zero time, Eq.~\eqref{eq:saddle_point_ent_asymm_gibbs}, in the previous inequality and the one at late times, Eq.~\eqref{eq:large_time_asymm_gibbs}, in the condition~\eqref{eq:condition_mpemba_t_I} for the QMPE, we conclude that these are satisfied if and only if
\begin{equation}\label{eq:cond_mpemba_gibbs}
\left\{\begin{array}{l}
\displaystyle g_{\beta}^{(n)}(\delta_1, h_1)> g_{\beta}^{(n)}(h_2, \delta_2),\\
\\
\displaystyle\int_{-k^*_{\zeta}}^{k^*_{\zeta}}\frac{{\rm d} k}{2\pi}\Upsilon_{\beta, k}^{(n)}(\delta_1, h_1)
<\int_{-k^*_{\zeta}}^{k^*_{\zeta}}\frac{{\rm d} k}{2\pi}\Upsilon_{\beta, k}^{(n)}(\delta_2, h_2),
\end{array}\right.
\end{equation}
for $t>t_I$. The first striking feature of this result is that, due to the intricate dependence of $g_{\beta}^{(n)}(\delta, h)$ and $\Upsilon_{\beta, k}^{(n)}(\delta, h)$ on the R\'enyi index $n$, the conditions in Eq.~\eqref{eq:cond_mpemba_gibbs} may be satisfied for some $n$ but not for others. To see this, we can take 
two XY spin chains whose ground states show QMPE, thus $\Delta S_A^{(n)}(\delta_1, h_1, \beta\to\infty, t)<\Delta S_A^{(n)}(\delta_2, h_2, \beta\to\infty, t)$ for $t>t_I$, and then obtain the initial temperature $\beta_{\rm M}$ for which their large time entanglement asymmetries equate, $\Delta S_A^{(n)}(\delta_1, h_1, \beta_{\rm M}, t)=\Delta S_A^{(n)}(\delta_2, h_2, \beta_{\rm M}, t)$. 

Solving that equation using the large time expression of Eq.~\eqref{eq:large_time_asymm_gibbs}, or by equating the second condition in Eq.~\eqref{eq:cond_mpemba_gibbs}, is generally difficult. A much easier way is by rewriting Eq.~\eqref{eq:cond_mpemba_gibbs} in a more explicit form as a function of the couplings $h$, $\delta$ and the inverse temperature $\beta$ of the initial chain. In fact, expanding in Eq.~\eqref{eq:large_time_asymm_gibbs} the velocity $v_k$ and $\Upsilon_{\beta, k}(\delta, h)$ around $k=0$ and $k^*_\zeta$ as $k^*_{\zeta}\simeq 1/(2\zeta)$, we obtain that the R\'enyi entanglement asymmetry decays 
as $\ell^4/t^3$ from any Gibbs ensemble $\rho_\beta$. In particular, 
for $n=2$,
\begin{multline}\label{eq:asymm_temp_asymp_n_2}
\Delta S_A^{(2)}(t)\simeq \frac{\ell \delta^2 }{384 (-1 + h^2)^2 \pi \zeta^3}\left[(1 + h)^2 
\right.\\ \left.\times\tanh^2((-1 + h)\beta)+ (-1 + 
      h)^2 \tanh^2((1 + h) \beta)\right].
\end{multline}
and, for $n=3$,
\begin{multline}\label{eq:asymm_temp_asymp_n_3}
\Delta S_A^{(3)}(t)\simeq \frac{\ell \delta^2}{512(1-h^2)^2\pi \zeta^3}
\left[2+2h^2\right.\\
\left.-\frac{(1+h)^2}{(1-2\cosh((-1+h)\beta))^2}
-\frac{(-1+h)^2}{(1-2\cosh((1+h)\beta))^2}\right].
\end{multline}
In the left and middle panels of Fig.~\ref{fig:beta_M}, we check Eq.~\eqref{eq:asymm_temp_asymp_n_2} against the exact numerical results.
Let us take now the XY spin chain along the curve $\delta^2+h^2=1$. As we already mentioned, this family of chains is peculiar because their ground states are the tilted ferromagnetic configurations, which have been the prototypical instance of initial pure states to investigate the QMPE theoretically~\cite{amc-23} and experimentally~\cite{joshi-24}. 
For $h\geq 0$, any pair of ground states in this curve shows Mpemba effect. Therefore, we 
can take $\delta_1=1$, $h_1=0$, whose ground state breaks the most the symmetry along the curve, and find the $\beta_M$ that solves the equation $\Delta S_A^{(n)}(1, 0, \beta_{\rm M}, t)=\Delta S_A^{(n)}(\delta, \sqrt{1-\delta^2}, \beta_{\rm M}, t)$ using Eq.~\eqref{eq:asymm_temp_asymp_n_2} for $n=2$ and Eq.~\eqref{eq:asymm_temp_asymp_n_3} for $n=3$.
In Fig.~\ref{fig:beta_M}, we represent the values of $\beta_{\rm M}$ that we obtain as a function of $h$. According to this plot, for temperatures $\beta>\beta_{\rm M}$ and fixed $n$, the entanglement asymmetries in a quench from $h_1=0$, $\delta_1=1$ and another chain with $\delta<1$ and $h=\sqrt{1-\delta^2}$ always cross and there is QMPE, while for $\beta<\beta_{\rm M}$, this phenomenon disappears. As we can see, $\beta_{\rm M}$ depends on the R\'enyi index $n$, implying that there exists an interval of $\beta$ for which the different R\'enyi entanglement asymmetries give an opposite answer as to whether the Mpemba effect occurs.  

We conclude this section by proving that the QMPE effect disappears when the temperature of the chains is very large for any pair of couplings $h, \delta$. In Eqs.~\eqref{eq:large_temp_gibbs_n_2} and~\eqref{eq:large_temp_gibbs_n_3}, we obtained the expression for the $n=2$ and $3$ R\'enyi entanglement asymmetries at large temperatures before the quench.
The late time behavior of the entanglement asymmetries for large temperatures can be derived by expanding Eqs.~\eqref{eq:asymm_temp_asymp_n_2} and \eqref{eq:asymm_temp_asymp_n_3} around $\beta=0$. We find 
\begin{equation}
\Delta S_A^{(2)}(\delta, h, \beta, t)\simeq \frac{\ell \delta^2 \beta^2}{192\pi\zeta^3}
\end{equation}
for $n=2$ and 
\begin{equation}
\Delta S_A^{(3)}(\delta, h, \beta, t)\simeq \frac{\ell \delta^2\beta^2}{128 \pi \zeta^3}
\end{equation}
when $n=3$. We observe that both at $t=0$ and $t\to\infty$ the entanglement asymmetries do not depend on the initial external magnetic field $h$ and they grow monotonically with the anisotropy parameter $\delta$. Therefore, in this case, the entanglement asymmetries of two different initial Gibbs ensembles $\rho_{\beta}(\delta_1, h_1)$ and $\rho_{\beta}(\delta_2, h_2)$ never satisfy the conditions of Eqs.~\eqref{eq:condition_mpemba_t_0} and \eqref{eq:condition_mpemba_t_I} for the occurrence of the QMPE.

\section{Conclusions}\label{sec:conclusions}

Quantum systems are commonly described as mixed states, especially
when they are subject to manipulations and interactions with the external environment. For instance, this is what happens in the observation of the QMPE in the trapped-ion quantum simulator studied in Ref.~\cite{joshi-24}.

For this reason, the goal of this manuscript has been to investigate how the mixedness of a state affects the symmetry restoration of a $U(1)$ symmetry and the occurrence of the QMPE. We have tackled this question by considering three distinct cases. In Sec.~\ref{sub:gainloss}, we have explored a scenario where an initially pure state evolves into a mixed state due to gain and loss dissipation. It turns out that, if gain and loss are balanced, they do not alter the conditions for the occurrence of the QMPE compared to the unitary case (see Fig.~\ref{fig:dissfirst}). The only difference is that the entanglement asymmetry decays exponentially in time, rather than algebraically. However, generic gain and loss terms can affect the occurrence of the QMPE as they shift the instant at which the crossing of the entanglement asymmetries happens. These findings are supported by numerical calculations as well as analytical predictions, derived by extending the quasiparticle picture to the weakly-dissipative hydrodynamic regime. One important 
difference with respect to the unitary evolution is that, in the presence of gain and loss dissipation, the stationary state of the subsystem does not depend on its initial state.
%the initial conditions. 
This is analogous to the classical Mpemba effect, in which the final equilibrium state is the same for all the initial conditions.
A very similar phenomenology is observed in the presence of local dephasing, i.e. local rotations of the spins around the $z$-axis (see Fig.~\ref{fig:ff_dephasing}). Indeed, the main lesson we learn from Sec.~\ref{sub:dephasing} is that this noise makes the entanglement asymmetry decrease and can shift the crossing time at which the QMPE occurs towards larger times as the dephasing rate increases, depending on the pair of initial states considered. To this end, we have also developed the equations of motion for the two-point functions in generic systems of free fermions with dephasing. The third scenario that we have considered is a global quantum quench from a system at a finite temperature, which is described by the Gibbs ensemble. In this case, the origin of the mixedness of the state is the configuration at time $t=0$, while the dynamics is purely unitary. In Sec.~\ref{sec:temperature}, we have first studied the entanglement asymmetry in the initial system and we have analyzed how it behaves between the ground state and the infinite temperature limit, where the symmetry is recovered. After the quench, we have observed that it exists a \textit{critical} temperature from which the QMPE disappears (see Fig.~\ref{fig:asymm_n_2_temp}). Even though its value depends on the specific R\'enyi index of the entanglement asymmetry, we expect that the critical temperature increases as $n$ increases.

There are several hints for future directions one could explore starting from our manuscript. For instance, a non-unitary dynamics can arise from local measurements performed during the time evolution, followed by post-selecting specific measurement outcomes. The action of this non-unitary evolution on the entanglement entropy has been studied in Ref.~\cite{turkeshi2023} for the XY model~\eqref{eq:Ham_XY} and it would be interesting to compare the effect of measurements with the one induced by dissipation. Unlike the entanglement entropy, which
is not a good measure of entanglement in mixed states, the entanglement asymmetry does quantify how much these states break a symmetry; nevertheless, it would be interesting to explore if one can define a probe of symmetry breaking based on entanglement measures of mixed states, such as entanglement negativity~\cite{vw-02, plenio-05}.
%Another possible future perspective concerns the introduction of a measure of symmetry breaking for mixed states, e.g. by using the entanglement negativity~\cite{vw-02, plenio-05}, which is commonly used to compute the entanglement in this kind of systems. 
An analytical challenge that our work leaves open is understanding why the charged moments defined in Eq.~\eqref{eq:Znalpha} do not factorize in the replica space, as happens in the absence of dissipation or at zero temperature. This prevents us from finding an analytical expression of the asymmetry for a generic index $n$ and taking the replica limit.
Finally, even though so far we have focused on free systems, a natural extension of our findings would be studying the entanglement asymmetry under a non-unitary dynamics or with an initial thermal state in interacting systems. 

\subsection*{Acknowledgments}
We thank Vincenzo Alba, Fabio Caceffo, Pasquale Calabrese and Colin Rylands for useful discussions and collaborations on related topics. FA acknowledges support from ERC under Consolidator Grant number 771536 (NEMO). SM thanks the support from the Caltech Institute for Quantum Information and Matter and the Walter Burke Institute for Theoretical Physics at Caltech.
VV acknowledge support from the French National
Research Agency via QUBITAF (ANR-22-PETQ-0004,
Plan France 2030).

\appendix
\section{Analytical details}
\subsection{Gain and loss}\label{app:sub1}
The goal of this appendix is to provide a step-by-step derivation of the result~\eqref{eq:charged_m} that describes the dynamics of the charged moments in the weakly dissipative regime.

Let us start by analyzing the behavior of the charge moments in the short and large time limits.
At time $t\to 0$, the symbol~\eqref{eq:symbol_dissipation} reads
\begin{equation}\label{eq:symbol_asymptotict0}
\begin{split}
\mathcal{G}(k,t\to 0)&=  \lambda(t)(\cos \Delta_k \sigma^z+\sin \Delta_k \sigma^y )\\&-\frac{\gamma_--\gamma_+}{\gamma_++\gamma_-}(1-\lambda(t))\sigma^z,
\end{split}
\end{equation}
where we keep the terms $e^{-(\gamma_++\gamma_-)t}$ because they are constant in the weakly-dissipative hydrodynamic limit.
As $t\to \infty$, the terms $e^{\pm 2it\epsilon_k}$ average to zero in Eq.~\eqref{eq:symbol_dissipation} and the symbol reduces to
\begin{equation}\label{eq:symbol_asymptotic}
\begin{split}
\mathcal{G}(k,t\to \infty)=  \lambda(t)\cos \Delta_k \sigma^z-\frac{\gamma_--\gamma_+}{\gamma_++\gamma_-}(1-\lambda(t))\sigma^z.
\end{split}
\end{equation}
We can now exploit in Eq.~\eqref{eq:numerics} the 
block Toeplitz structure of the two-point fermionic correlation function to compute the charged moments in these two limits. From the results of Ref.~\cite{amvc-23}, we know that, if $T_\ell[g]$ is a $(2\ell )\times (2\ell )$ dimensional block Toeplitz matrix with symbol a $2\times 2$ matrix $g$, then for large $\ell$
\begin{equation}\label{eq:conj_1}
\det\left(I+\prod_{j=1}^nT_\ell[g_j] T_\ell[g_j']^{-1}\right)\sim e^{\ell A},
\end{equation}
where the exponent $A$ is given by
\begin{equation}\label{eq:conj2}
A=\int_{0}^{2\pi}\frac{{\rm d}k}{2\pi}\log\det\left[I+\prod_{j=1}^n g_j(k)g_j'(k)^{-1}\right].
\end{equation}
Employing this formula in Eq.~\eqref{eq:numerics} with the symbol~\eqref{eq:symbol_asymptotic}, we derive the stationary value of $Z_n(\boldsymbol{\alpha},t)$ at large times,
\begin{equation}\label{eq:stat1}
 \log Z_n(\boldsymbol{\alpha}, t\to\infty)\sim 
 \frac{\ell}{2}\int_0^{2\pi} \frac{{\rm d}k}{2\pi}\log \det \mathcal{M}_{\boldsymbol{\alpha}}^{(n)}(k,t\to\infty),
\end{equation}
where $\mathcal{M}_{\boldsymbol{\alpha}}^{(n)}(k, t)$ was defined in Eq.~\eqref{eq:M}. We observe that, in the large time limit, the symbol in Eq.~\eqref{eq:symbol_asymptotic} commutes with $\sigma^z$ and the dependence on $\alpha_{j,j+1}$ in $\mathcal{M}_{\boldsymbol{\alpha}}^{(n)}(k, t\to\infty)$ simplifies. 
In fact, after some algebra, Eq.~\eqref{eq:stat1} becomes
\begin{equation}\label{eq:stat2}
   \log Z_n(\boldsymbol{\alpha}, t\to\infty)\sim \ell\int_0^{2\pi} \frac{{\rm d}k}{2\pi}h_n(n(k, t)),
\end{equation}
with   
$h_n(x)=\log\left[x^n+(1-x)^n\right]$
and 
\begin{equation}\label{eq:occupation}
\begin{split}
    n(k, t)=\frac{1}{2}\left[1-\lambda(t)\cos \Delta_k  +\frac{\gamma_--\gamma_+}{\gamma_++\gamma_-}(1-\lambda(t))\right]
\end{split}
\end{equation}
is the density of occupied modes with momentum $k$~\cite{alba2022hydrodynamics}. According to Eq.~\eqref{eq:stat2}, in the large $t$ limit, the charged moments do not depend on $\alpha_{j,j+1}$. This result implies that $\Delta S_A^{(n)}(t)\to 0$ when $t\to \infty$, indicating that the particle number symmetry is restored in $A$ in the stationary state.

Following the same reasoning, we compute the charged moments in the limit $t\to 0$, by applying in Eq.~\eqref{eq:numerics} the formula~\eqref{eq:conj_1} with the symbol~\eqref{eq:symbol_asymptotict0}. We get
\begin{equation}\label{eq:init_time}
 \log Z_n(\boldsymbol{\alpha}, t\to0)\sim \frac{\ell}{2}\int_{0}^{2\pi}
 \frac{{\rm d}k}{2\pi}\log\det \mathcal{M}_{\boldsymbol{\alpha}}^{(n)}(k,t\to0).
\end{equation}

We can now determine the contribution of the quasiparticles created in the quench to the charged moments from the difference between their short and long time limits,

\begin{multline}\label{eq:diff_short_large_time}
\log\frac{Z_n(\boldsymbol{\alpha},t\to\infty)}{Z_n(\boldsymbol{\alpha}, t\to0)}\sim
 \log Z_n(\boldsymbol{0}, t\to\infty)\\
 -\frac{\ell}{2}\int_{0}^{2\pi}
 \frac{{\rm d}k}{2\pi}\log\det\mathcal{M}_{\boldsymbol{\alpha}}^{(n)}(k,t\to0),
\end{multline}
where we took into account that $Z_n(\boldsymbol{\alpha}, t\to \infty)\sim Z_n(\boldsymbol{0}, t\to\infty)$.
Within the quasiparticle picture, the factor $\ell$ in the right-hand 
side of this result should be interpreted as the number of entangled 
pairs shared by $A$ and  $B$ at $t\to\infty$. Consequently,
this expression can be extended to any time step $t$ by taking into account the number of entangled
excitations between $A$ and $B$ at each moment. Assuming that the quasiparticle velocity $v_k$ is not affected by the dissipation, this can be done by inserting the
function min$(2\zeta|v_k|, 1)$ in the momentum integrals of the right-hand side of Eq.~\eqref{eq:diff_short_large_time}.
We can then conclude that, in the weakly-dissipative hydrodynamic regime, the charged moments evolve as in Eq.~\eqref{eq:charged_m},
\begin{equation}
Z_n(\boldsymbol{\alpha}, t)= Z_n(\boldsymbol{0},t)e^{B_n(\boldsymbol{\alpha}, \zeta)\ell},
\end{equation}
where 
\begin{equation}
 B_n(\boldsymbol{\alpha}, \zeta)=\int_0^{2\pi} \frac{{\rm d}k}{4\pi}x_k(\zeta)\log \left[\frac{\det \mathcal{M}_{\boldsymbol{\alpha}}^{(n)}(k, t\to0)}{\det\mathcal{M}_{\boldsymbol{0}}^{(n)}(k, t\to 0)}\right]
\end{equation}
and
%\bigskip
\begin{multline}\label{eq:neutral_mom_diss}
Z_n(\boldsymbol{0},t)=\ell\int_0^{2\pi} \frac{{\rm d}k}{2\pi}\mathrm{min}(2\zeta |v_k|,1)h_n(n(k)) \\
+\frac{\ell}{2}\int_{0}^{2\pi}\frac{{\rm d}k}{2\pi}x_k(\zeta)\log \det\mathcal{M}_{\boldsymbol{0}}^{(n)}(k, t\to 0).
 \end{multline}
Eq.~\eqref{eq:neutral_mom_diss} is the time evolution of the (neutral) moments of $\rho_A$ in the presence of linear gain and loss dissipation found in Ref.~\cite{alba2021spreading}. 

\subsection{Local dephasing}\label{app:sub2}

In this appendix, we derive the equations of motion for \( G_{mn}=\langle a^{\dagger}_ma_n \rangle \) and \( F_{mn}=\langle a^{\dagger}_ma^{\dagger}_n \rangle \) in the presence of local dephasing. We show that they obey a linear first-order differential equation, which makes them easy to be computed numerically.

Our starting point is the introduction of \( 2N \) Majorana fermions, defined as
\begin{equation}
c_{2m-1}=a_m+a_m^\dag, \quad c_{2m}=i(a_m-a_m^\dag),
\end{equation}
which satisfy the anticommutation relations \( \{ c_k , c_l \} =2 \delta_{k,l} \) for all \( k,l=1,\dots, 2N \).
In terms of them, the most general quadratic Hamiltonian and Lindblad operators are of the form
\eq{
H=\frac{i}{4}\sum_{k,l=1}^{2N}H_{kl} c_k c_l \, , \qquad
L_\alpha=\frac{i}{4}\sum_{k,l=1}^{2N}L_{\alpha,kl} c_k c_l,
\label{eq:hlmatrix}}
where  the condition $H_{lk}=-H_{kl}$ is required to be the Hamiltonian $H$ Hermitian. 

Furthermore, it is convenient to introduce ordered strings of Majorana operators
\eq{\Gamma_{\underline{\nu}}=c_1^{\nu_1} \dots c_{2N}^{\nu_{2N}}, \qquad
\nu_i \in \left\{ 0,1 \right\}
\label{eq:npf}}
where $\underline{\nu} = (\nu_1,\dots,\nu_{2N})$ denotes whether the corresponding Majorana operator $c_i$ is present in the string $\Gamma_{\underline{\nu}}$.
Finally, one can define \emph{superoperators} $\hat{c}_j$, $\hat{c}_j^\dagger$ acting on these strings that create or annihilate Majorana operators at position $j$ as
\eq{
\begin{split}
\hat{c}_j \Gamma_{\underline{\nu}} = &\delta_{1,\nu_j}\pi_j\Gamma_{\underline{\nu}'}, \quad
\hat{c}_j^{\dag} \Gamma_{\underline{\nu}} = \delta_{0,\nu_j}\pi_j\Gamma_{\underline{\nu}'}, \quad\\
&\nu'_i = \left\{
\begin{array}{ll}
1-\nu_i, & i=j \\
\nu_i, & i \ne j
\end{array} \right.
\end{split}
\label{eq:so}}
The sign factor $\pi_j= \exp\left( i\pi \sum_{k=1}^{j-1}\nu_k\right)$ is introduced to guarantee that they verify the canonical anticommutation
relations $\{ \hat{c}_i , \hat{c}_j^{\dag} \} = \delta_{i,j}$ and $\left\{ \hat{c}_i , \hat{c}_j \right\}=0$.

As already seen in Refs.~\cite{prosen2008, Eislerdephasing2011}, using these superoperators, the Liouvillian defined in Eq.~\eqref{eq:lind} can be rewritten as
\eq{
\mathcal{L^\dag} = -\sum_{k,l=1}^{2N} \tilde H_{kl} \hat{c}^{\dag}_k \hat{c}_l
+ \frac{1}{2} \sum_{i,j,k,l=1}^{2N} \sum_{\alpha}
L^T_{\alpha,ij} L_{\alpha,kl} \hat{c}^\dag_i\hat{c}^{\dag}_k \hat{c}_j\hat{c}_l 
\label{eq:liou}}
provided the Lindblad operators satisfy $L_\alpha^\dag = L_\alpha$
for all $\alpha$. In this expression, $\tilde H_{kl}$ are the entries of the matrix ${\bf \tilde H}={\bf H}+\frac{1}{2}\sum_\alpha{\bf L}^T_\alpha {\bf L}_\alpha$, that includes the 
Hamiltonian evolution, in our case given by Eq.~\eqref{eq:Ham_XY} with $\delta=0$, $h=0$, and a damping term, responsible for the decay of the $n$-point functions.
This part, which is of the same form as in the case of linear gain and loss dissipation, is quadratic and respects the Gaussianity of the initial state. However, the second term in Eq.~\eqref{eq:liou} is quartic in the
superoperators and, consequently, the time-evolved density matrix $\rho(t)$ is not Gaussian. In general, the Liouvillian~\eqref{eq:liou} cannot be diagonalized, although in the case of local dephasing it further simplifies, allowing us to calculate the evolution of the two-point 
correlators efficiently.

The ingredient that we need for deriving the equations of motion for them is computing the action of the Linbladian~\eqref{eq:liou} on any string of Majorana operators $\mathcal{L}(\Gamma_{\underline{\nu}})$.
Following the prescriptions in Ref.~\cite{Eislerdephasing2011}, a simple calculation yields to the following Liouvillian
\eq{
\begin{aligned}
&\mathcal{L}_{a} =
\sum_{m} \left[ -i \left( \hat{a}_m^\dag \hat{a}_{m+1}  + \hat{a}_{m+1}^\dag \hat{a}_{m} \right) -\frac{\gamma_d}{2} \hat{a}_m^\dag \hat{a}_{m} \right],
\\
&\mathcal{L}_{b} = \sum_{m} \left[ i \left( \hat{b}_m^\dag \hat{b}_{m+1} + \hat{b}_{m+1}^\dag \hat{b}_{m} \right) -\frac{\gamma_d}{2} \hat{b}_m^\dag \hat{b}_{m}  \right],
\\
&\mathcal{L}_{ab} = -\sum_{m} \gamma_d \hat{b}_{m}^\dag \hat{b}_{m} \hat{a}_{m}^\dag \hat{a}_{m},
\end{aligned}
\label{eq:liouqep}}
where we have introduced the canonical transformation for the superoperators:
\begin{equation}
    \begin{aligned}
        &\hat{a}_m = \frac{1}{\sqrt{2}}\left(\hat{c}_{2m-1} - i\hat{c}_{2m}\right),\\ 
&\hat{b}_m = \frac{1}{\sqrt{2}}\left(\hat{c}_{2m-1} + i\hat{c}_{2m}\right).
    \end{aligned}
\end{equation}

Then one needs to evaluate the action of all the elements in Eq.~\eqref{eq:liouqep} on $a_m^{\dagger}a_m$ and $a_m^{\dagger}a^{\dagger}_n$.
Observing that
\begin{equation}
\begin{aligned}
&\hat{a}_m a^{\dagger}_k=\frac{1}{\sqrt{2}}\delta_{mk}, \qquad \hat{a}_ma_k=0, \qquad \hat{a}^{\dagger}_m=\sqrt{2}a^{\dagger}_m, \\
& \hat{b}_m a^{\dagger}_k=0, \qquad \hat{b}_ma_k=\frac{1}{\sqrt{2}}\delta_{mk}, \qquad \hat{b}^{\dagger}_m=\sqrt{2}a_m,
\end{aligned}
\end{equation}
one gets
\begin{equation}
    \begin{aligned}
        \frac{\dd}{\dd t} G_{m,n} =& -i 
\left( G_{m-1,n} + G_{m+1,n} - G_{m,n-1} - G_{m,n+1} \right) \\&- \gamma_d\left( G_{m,n}-\delta_{mn} G_{m,n}\right),
\label{eq:evolg}
    \end{aligned}
\end{equation}
and similarly
\eq{
\frac{\dd}{\dd t} F_{m,n} = -i \left( F_{m-1,n} + F_{m+1,n} \right) - \frac{\gamma_d}{2} F_{m,n}, \\
\label{eq:evolf}}
where we have used $\frac{\dd}{\dd t}G_{m,n}=\langle \mathcal{L}_d\left(a_m^{\dagger}a_m\right)\rangle$ and $\frac{\dd}{\dd t}F_{m,n}=\langle \mathcal{L}_d\left(a_m^{\dagger}a^{\dagger}_m\right)\rangle$.

%\bibliography{biblio.bib}

\end{document}